\documentclass[final]{agujournal2019}
\usepackage{url} 
\usepackage{lineno}
\usepackage{amsmath,amsfonts,amssymb}
\usepackage[inline]{trackchanges} 
\usepackage{soul}
\usepackage[normalem]{ulem}
\usepackage{xr}
\usepackage{xcolor, amsmath}

\usepackage{scalerel,stackengine}
\stackMath
\newcommand\reallywidehat[1]{%
\savestack{\tmpbox}{\stretchto{%
  \scaleto{%
    \scalerel*[\widthof{\ensuremath{#1}}]{\kern.1pt\mathchar"0362\kern.1pt}%
    {\rule{0ex}{\textheight}}
  }{\textheight}%
}{2.4ex}}%
\stackon[-6.9pt]{#1}{\tmpbox}%
}

\newcommand*\patchAmsMathEnvironmentForLineno[1]{
  \expandafter\let\csname old#1\expandafter\endcsname\csname #1\endcsname
  \expandafter\let\csname oldend#1\expandafter\endcsname\csname end#1\endcsname
  \renewenvironment{#1}
  {\linenomath\csname old#1\endcsname}
  {\csname oldend#1\endcsname\endlinenomath}}
  \newcommand*\patchBothAmsMathEnvironmentsForLineno[1]{
  \patchAmsMathEnvironmentForLineno{#1}
  \patchAmsMathEnvironmentForLineno{#1*}}
  \AtBeginDocument{
  \patchBothAmsMathEnvironmentsForLineno{equation}
  \patchBothAmsMathEnvironmentsForLineno{align}
  \patchBothAmsMathEnvironmentsForLineno{flalign}
  \patchBothAmsMathEnvironmentsForLineno{alignat}
  \patchBothAmsMathEnvironmentsForLineno{gather}
  \patchBothAmsMathEnvironmentsForLineno{multline}
}

\newcommand{\mycomment}[1]{}

\usepackage{color}

\draftfalse

\journalname{Journal of Advances in Modeling Earth Systems (JAMES)}

\begin{document}

%
%


\title{A Data-Driven Approach for Parameterizing Ocean Submesoscale  Buoyancy Fluxes}

\authors{Abigail Bodner\affil{1}, Dhruv Balwada\affil{2}, and Laure Zanna\affil{1,3}}

\affiliation{1}{Courant Institute of Mathematical Sciences, New York University, New York, NY, USA}
\affiliation{2}{Lamont-Doherty Earth Observatory, Columbia University, Palisades, NY, USA}
\affiliation{3}{Center for Data Science, New York University, New York, NY, USA}




\correspondingauthor{A. Bodner}{abodner@mit.edu}




\begin{keypoints}
\item A data-driven parameterization for mixed layer vertical buoyancy fluxes is developed using a Convolutional Neural Network (CNN).
\item The CNN demonstrates high offline skill over a wide range of dynamical regimes and filter scales.
\item We identify strong dependency on the large scale strain field, currently missing from oceanic submesoscale parameterizations.
\end{keypoints}

%
%

%
%


\begin{abstract}
Parameterizations of $O(1-10)$km submesoscale flows in General Circulation Models (GCMs) represent the effects of unresolved  vertical buoyancy fluxes in the ocean mixed layer. These submesoscale flows interact non-linearly with mesoscale and boundary layer turbulence, and it is challenging to account for all the relevant processes in physics-based parameterizations. In this work, we present a data-driven approach for the submesoscale parameterization, that relies on a Convolutional Neural Network (CNN) trained to predict mixed layer vertical buoyancy fluxes  as a function of relevant large-scale variables. The data used for training is given from 12 regions sampled from the global high-resolution MITgcm-LLC4320 simulation. When compared with the baseline of a submesoscale physics-based parameterization, the CNN demonstrates high offline skill across all regions, seasons, and filter scales tested in this study. 
During seasons when submesoscales are most active, which generally corresponds to winter and spring months, we find that the CNN prediction skill tends to be lower than in summer months. The CNN exhibits strong dependency on the mixed layer depth and on the large scale strain field, a variable closely related to frontogenesis, which is currently missing from the submesoscale parameterizations in GCMs.

\end{abstract}

\section*{Plain Language Summary}
Upper ocean turbulence is an important control on energy and heat exchanges between the atmosphere and ocean systems, and usually manifests at spatial scales that are too small to be resolved in climate models. Parameterizations are often used to estimate missing physics using information from the resolved flow. In contrast to traditional approaches, this work uses machine learning to predict unresolved properties of upper ocean turbulence. This new approach demonstrates high performance over a range of scales, locations and seasons, with potential to help reduce climate model biases.

\section{Introduction}
General Circulation Models (GCMs) and future climate change projections are notoriously sensitive to parameterizations of unresolved phenomena at the ocean-atmosphere interface \cite{SROCC, RN1}. Of particular importance is the ocean mixed layer, where turbulence modulates the transfer of properties – such as heat, momentum, and carbon – between the atmosphere and ocean interior \cite<e.g.,>[]{frankignoul1977stochastic,bopp2015pathways,su2020high}. Turbulence in the ocean mixed layer spans a wide range of scales, from the $O(100)$km mesoscales, to $O(1-10)$km submesoscales, to $O(1-100)$m boundary layer turbulence, and all the way down to the molecular scales. A sensitive dynamical interplay between turbulence across all relevant scales sets the stratification in the upper ocean \cite{treguier2023mixed}. As opposed to mixing and homogenization dominated by boundary layer turbulence, submesoscale flows play a particularly important role in contributing to vertical transport in the ocean mixed layer primarily by shoaling, or restratifying, the mixed layer \cite{boccaletti2007mixed,mcwilliams2016submesoscale, taylor2023submesoscale}.  

The restratification effect is at leading order a result of instabilities formed along mixed layer fronts-- composed of sharp density gradients and vertically oriented isopycnals \cite{fox2008parameterization,gula2022submesoscale}. One of the primary submesoscale instabilities, known as mixed layer instabilities, produces vertical buoyancy fluxes (VBF) by slumping the fronts and restratifying the mixed layer. The effect of submesoscale restratification cannot be resolved in many GCMs, and is currently parameterized by the  Mixed Layer Eddy parameterization \cite<hereafter MLE,>[]{fox2011parameterization}. 
Recent advances in submesoscale parameterization development propose new relationships between MLE and large scale properties of the mesoscale field \cite<e.g.,>[]{zhang2023parameterizing} and boundary layer turbulence \cite<e.g.,>[]{bodner2023modifying}, 
yet these new approaches still struggle to capture the full range of complexity \cite{lapeyre2006oceanic,mahadevan2010rapid, bachman2017parameterization, callies2018baroclinic,ajayi2021diagnosing}.

Data-driven methods are emerging as powerful tools, with the ability to capture highly complex relationships between variables in turbulent flows.  Advances in machine learning based parameterizations have yielded promising results for subgrid closures such as for ocean mesoscale momentum fluxes \cite{bolton2019applications,zanna2020data,guillaumin2021stochastic,perezhogin2023generative}, ocean boundary layer mixing \cite{souza2020uncertainty,sane2023parameterizing}, 
and atmospheric boundary layer mixing \cite<e.g.,>[]{yuval2020stable, wang2022non,shamekh2023implicit}.
Numerous examples for other machine learning applications exist both in the atmosphere and the ocean for inference of flow patterns and structures from data \cite<e.g.,>[]{chattopadhyay2020predicting, dagon2022machine, xiao2023reconstruction, zhu2023deep}.

Here, we introduce a data-driven approach for parameterizing submesoscale-induced VBF in the ocean mixed layer. We train a Convolutional Neural Network (CNN) using high-resolution simulation data, with the goal of learning an improved functional relationship between mixed layer VBF and the large-scale variables that help set it. The data used to train and test the CNN is sampled from the MITgcm-LLC4320 ocean model \cite<hereafter LLC4320,>{menemenlis2021pre}, which simulated the global ocean at a resolution of $1/48^o$. 
The LLC4320 output has been widely studied for submesoscale applications, which cumulatively have demonstrated that submesoscale energetics and dynamics are captured relatively well down to its effective resolution \cite<e.g.,>[]{rocha2016mesoscale,  su2018ocean, gallmeier2023evaluation}. In this paper, we describe the processing of the LLC4320 data and CNN architecture in section \ref{sec:data}. Results and sensitivity tests of the CNN prediction on unseen data are presented and compared with the baseline of the MLE parameterization in section \ref{sec:results}. In section \ref{sec:features}, we apply two complimentary  methods to explain the  relationship learned by the CNN and the mixed layer VBF. Discussion and concluding remarks are given in section \ref{sec:conclusion}.

\section{Data and methods}
\label{sec:data}
\subsection{Processing the LLC4320}

The LLC4320 is a $1/48^o$ Massachusetts Institute of Technology general circulation model
(MITgcm), named after its Latitude‐Longitude polar Cap (LLC) grid with 4320 points on each of the 13 tiles. The LLC4320 is initialized from the Estimating the Circulation and Climate of the Ocean (ECCO), Phase II project, and is forced at the surface by atmospheric reanalysis, at 6 hourly temporal resolution. 
Model output from a total of 14 months is available at hourly frequency from September 2011 to  November 2012 \cite{forget2015ecco,menemenlis2008ecco2,menemenlis2021pre}. Before computing the CNN input and output variables, a common processing procedure was applied to the LLC4320 data, as described below.

Since the primary goal of our work is to parameterize the impact of submesoscale processes on mixed layer stratification, we focus on diagnosing subgrid VBF \cite{fox2008parameterization}. The large-scale fields, which may be resolved in a coarse-simulation, and subgrid impacts, which need to be parameterized, are defined with the help of filters. This is similar to approaches commonly used in the large-eddy simulation literature \cite{sagaut2005large}.
First, to reduce the data volume, we averaged all LLC4320 variables over periods of 12 hours, effectively time-filtering the fastest motions. Primarily composed of tides and internal waves, these fast-varying motions  have minimal impact on the VBF and so the filtering operation does not account them  \cite{balwada2018submesoscale, uchida2019contribution}. 
Next, a top-hat or coarsening spatial filter (denoted by $\overline{ \ \cdot \ }  $) was applied to decompose the simulation variables into large-scale and subgrid components. This allows us to define the subgrid VBF ($\overline{F^V}$) as:
\begin{equation}
    \overline{F^V} = \overline{w}\overline{b}-\overline{wb},
\end{equation}
where $\overline{w}$ and $\overline{b}$ correspond to the large-scale variables that can be resolved on a coarse-grid, and $\overline{wb}$ is the coarsened flux that is resolved in a high-resolution simulation.
We used filter scales of $1^o,{1/2}^o,{1/4}^o,{1/8}^o,{1/12}^o$, which are defined by the width of  the coarsening box, applied by averaging over a fixed number of grid points in the original LLC4320 grid.  As an example for the ${1/4}^o$ spatial filter, averages are taken over $12\times12$ grid points of the ${1/48}^o$ simulation grid.  

For simplicity, we restrict our approach to depth-averaged mixed layer properties, and to this end, all 3D variables are averaged over the mixed layer depth (denoted by superscript $z$ hereafter).
This also remains close to the formulation of the MLE parameterization \cite{fox2011parameterization}, where the parameterization is composed of a depth-independent amplitude and a vertical structure function that determines the shape of the parameterization over the mixed layer (Eq.~\eqref{eq:structure function} in the supplementary material). Here, the mixed layer depth,  H$_{ML}$, is defined as the depth at which the potential density anomaly, $\sigma_0$, increased by $0.03$ kg\  m$^{-3}$ from its value at 10m depth \cite{de2004mixed}. $\sigma_0$ is computed from the LLC4320 outputs of potential temperature and salinity fields, with reference pressure of 0 dbar and $\rho_0 = 1000$ kg\ m$^{-3}$.

\begin{figure}[ht!]
 \centering   \makebox[\textwidth]{\includegraphics[width=1\textwidth]{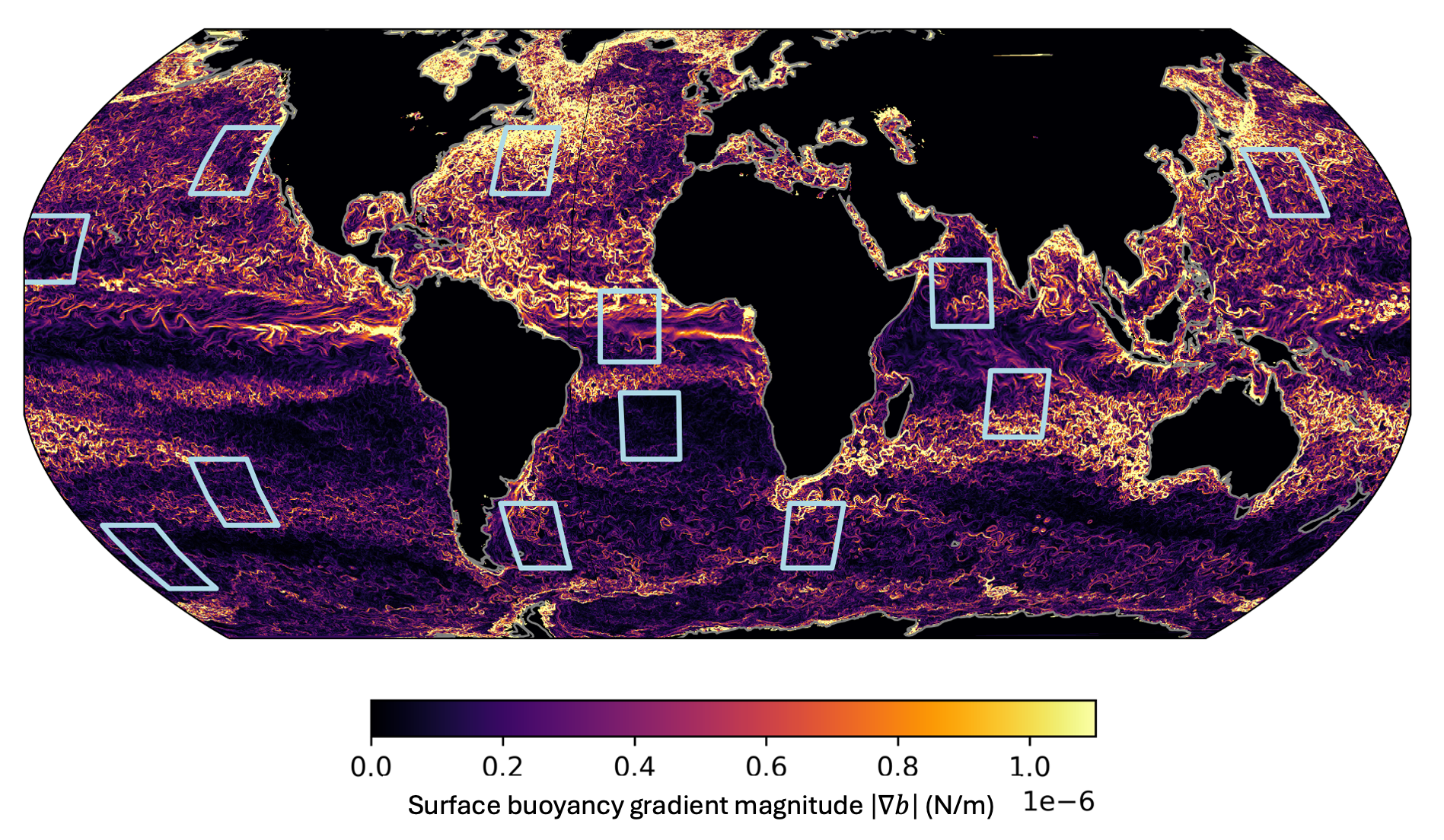}}%
    \caption{Snapshot of surface horizontal buoyancy gradient magnitude $|\nabla b|$ (N/m) given by the global LLC4320 simulation coarsened to $1/4^o$. Buoyancy gradients are a key contributor to the characteristics of submesoscale flows, and their properties vary significantly between regions and season (Fig. \ref{fig: wb cospectrum}), motivating the choice of sampled regions used in this study (light blue boxes, exact coordinates listed in Tab. \ref{tab:regions}).}
  \label{fig: grad b}
\end{figure}

To ensure that the training data represented a diverse set of dynamics, we included a mix of regions with strong and weak variability \cite<e.g.,>[]{torres2018partitioning} as shown by blue boxes in Fig. \ref{fig: grad b} (exact coordinates provided in Tab. \ref{tab:regions}). 
In the following section we describe how the LLC4320 data is  processed for each of these regions to diagnose the vertical buoyancy flux (CNN output) and a variety of inputs (Tab. \ref{tab:features})
in preparation for the CNN training. 

\subsection{Input and output features}
\label{subsec:input output}

\begin{figure}[ht!]
 \centering
   \makebox[\textwidth]{\includegraphics[width=1\textwidth]{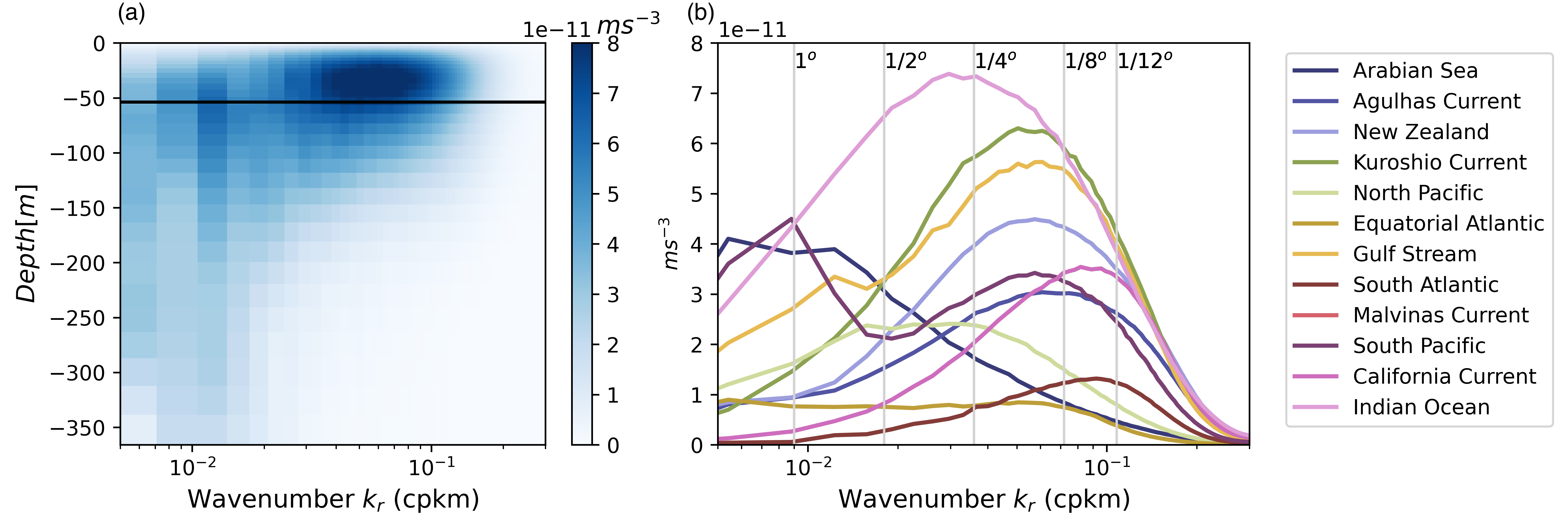}}%
    \caption{Isotropic cross-spectrum of  $w$ and $b$ in variance-preserving form, averaged over the entire LLC4230 simulation duration (14 months): (a) Example of the depth varying cross-spectrum in the Gulf Stream region , illustrating that the VBF is concentrated in the small scales and within the average mixed layer (black horizontal line). (b) Cross-spectrum averaged over the mixed layer depth for all regions. Vertical gray lines mark the filter scales used in this study with respect to the cross-spectrum variability. }
  \label{fig: wb cospectrum}
\end{figure}

The subgrid quantity we are parameterizing using the CNN is the depth-averaged mixed layer VBF ($\overline{F^V}^z$). This quantity is formally denoted as, 
\begin{equation}
\mathbb{Y}_{wb}:= \overline{F^V}^z = \overline{w}^z\overline{b}^z-\overline{wb}^z.
\end{equation}
VBF in the mixed layer is largely a result of submesoscale flows \cite{boccaletti2007mixed}.
This can be seen in the maximum cross-spectrum of $w$ and $b$, analogous to maximum VBF, which is found to be predominantly in the submesoscale range and confined to the mixed layer (Fig. \ref{fig: wb cospectrum}a). 
However, variability across scales can defer between the different regions, and the filter scale choice (illustrated by the grey lines in Fig. \ref{fig: wb cospectrum}b) will impact the properties of the large scale CNN inputs and subgrid flux output. We have included all regions to gain a variety of dynamical regimes in our training data, but test the performance of the CNN over the different filter scales, and selected unseen regions and parts of the timeseries in section \ref{sec:results}.

We choose to include input features that are correlated (Fig. \ref{fig: corr map}), or have known analytical relationships, with submesoscale VBF. We leverage the physical relevance demonstrated by variables that appear in the MLE parameterization \cite{fox2011parameterization,bodner2023modifying}, as well as correlated large-scale velocity derivatives \cite<e.g.,>[]{barkan2019role, balwada2021vertical,zhang2023parameterizing}. The input features (Tab. \ref{tab:features}) consist of the depth-averaged horizontal buoyancy gradient magnitude, $|\overline{\nabla b}^z|$, where buoyancy is defined as $b=-g \sigma_0/\rho_0$, and $g=9.81$ m/s$^2$ is the gravity acceleration; Coriolis parameter, $\overline{f}$; mixed layer depth, $\overline{H_{ML}}$; surface heat flux ,$\overline{Q}$; surface wind stress magnitude, ${|\overline{\tau}|} = \sqrt{\overline{\tau_x}^2+\overline{\tau_y}^2}$; boundary layer depth, $\overline{H_{BL}}$, an output of the LLC4320 computed from the Richardson number criteria with the critical value of 0.3 \cite< K‐profile Parameterization,>[]{large1994oceanic}; depth-averaged strain magnitude, $\overline{\sigma}^z = \sqrt{ ( \overline{u_x}^z-\overline{v_y}^z)^2 +( \overline{v_x}^z+\overline{u_y}^z)^2} $;
depth-averaged vertical vorticity, $\overline{\zeta}^z = \overline{v_x}^z-\overline{u_y}^z $; depth-averaged horizontal divergence, $\overline{\delta}^z = \overline{u_x}^z+\overline{v_y}^z$. Note that velocities $(u,v,w)$ and wind stresses $(\tau_x, \tau_y)$ are all interpolated to collocate with the tracer grid. 

\begin{figure}[ht!]
 \centering   \includegraphics[width=.7\textwidth]{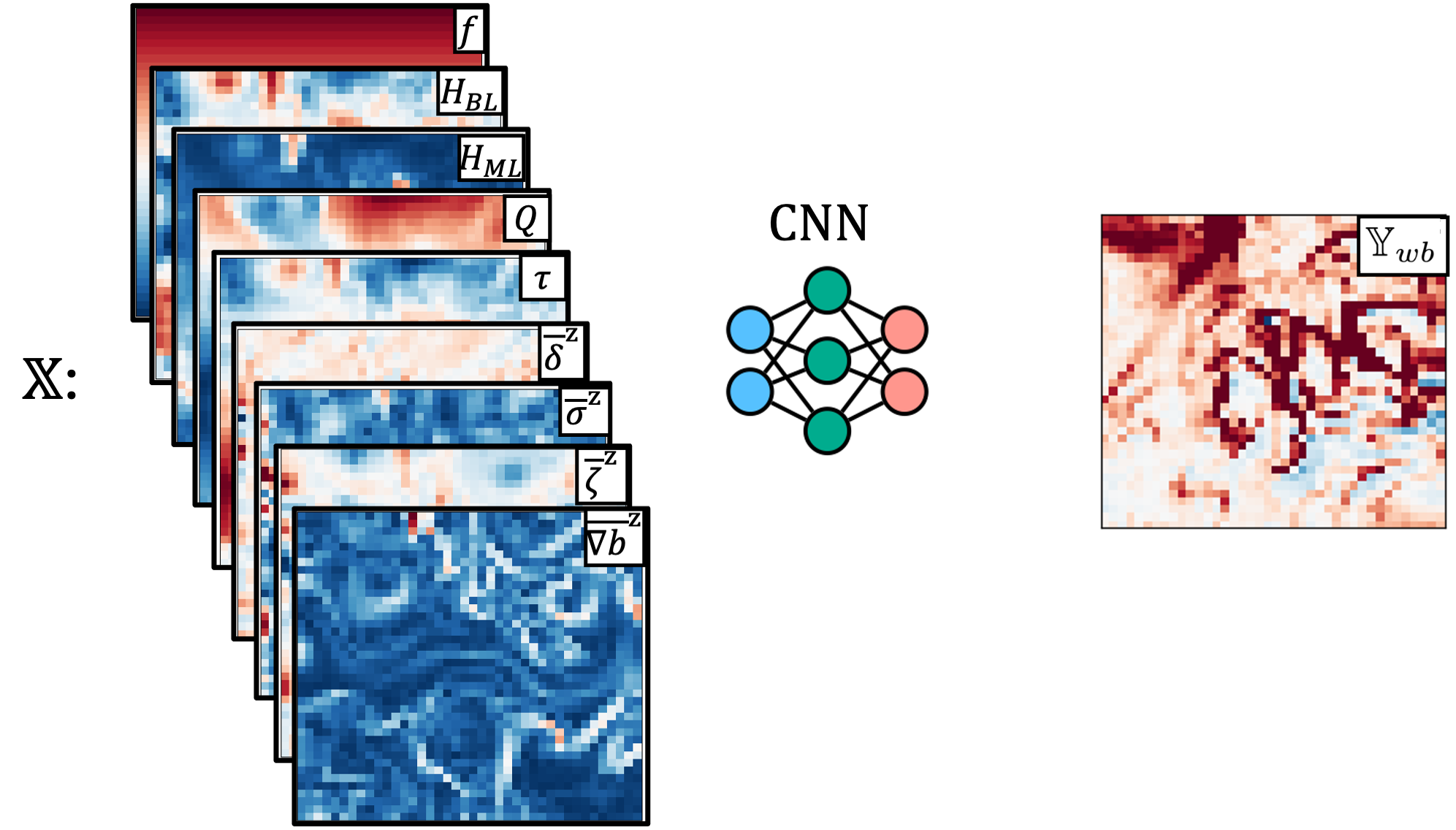}
    \caption{Schematic of the $1/4^o$ CNN method with 9 input features, $\mathbb{X}$, and one output, $\mathbb{Y}_{wb}$ (corresponding to Tab. \ref{tab:features}). The CNN architecture is described in section \ref{sec:NN}.}
  \label{fig: schematic}
\end{figure}

Formally, we define our 9 input features as,
\begin{equation}
    \mathbb{X} := (|\overline{\nabla b}^z|, \overline{f},\overline{H_{ML}}, \overline{Q}, {|\overline{\tau}|} ,\overline{H_{BL}},\overline{\sigma}^z,\overline{\zeta}^z, \overline{\delta}^z),
\end{equation}
and a single output as,
\begin{equation}
    \mathbb{Y}_{wb} :=\overline{w}^z\overline{b}^z-\overline{wb}^z
\end{equation}  
The CNN provides a prediction of the subgrid fluxes as a function of the large scale variables, such that $S (\mathbb{X})=\hat{\mathbb{Y}}_{wb}$ and $\hat{\mathbb{Y}}_{wb}\rightarrow \mathbb{Y}_{wb}$, where $S$ represents the CNN and $\hat{\mathbb{Y}}_{wb}$ its prediction.
Fig. \ref{fig: schematic} illustrates a schematic of the CNN with 9 input features and one output. 

\begin{table}[h]
    \centering
\begin{tabular}{l l }
\textbf{CNN inputs, $\mathbb{X}$} & \\
    Depth-averaged buoyancy gradient magnitude &
     $|\overline{\nabla b}^z| = \sqrt{(\overline{b_x}^z)^2+(\overline{b_y}^z)^2}$ \\
     Coriolis parameter & $\overline{f}$ \\
     Mixed layer depth & $\overline{H_{ML}}$ \\
     Surface heat flux & $\overline{Q}$ \\
     Surface wind stress magnitude & $|\overline{\tau}| = \sqrt{\overline{\tau_x}^2+\overline{\tau_y}^2}$  \\
     Boundary layer depth & $\overline{H_{BL}}$  \\
     Depth-averaged strain magnitude & $\overline{\sigma}^z = \sqrt{ ( \overline{u_x}^z-\overline{v_y}^z)^2 +( \overline{v_x}^z+\overline{u_y}^z)^2} $ \\
     Depth-averaged vertical vorticity  & $\overline{\zeta}^z = \overline{v_x}^z-\overline{u_y}^z $  \\
          Depth-averaged horizontal divergence & 
$\overline{\delta}^z = \overline{u_x}^z+\overline{v_y}^z$\\
&\\
\textbf{CNN Output, $\mathbb{Y}_{wb}$} & \\
Depth-averaged subgrid vertical buoyancy flux & $\mathbb{Y}_{wb} :=\overline{w}^z\overline{b}^z-\overline{wb}^z$\\

\end{tabular} 
    \caption{Input ($\mathbb{X}$) and output ($\mathbb{Y}_{wb}$) features used in the CNN method. Overbar represents the top-hat spatial filter and superscript $z$ represents a depth averaging over the mixed layer depth applied as part of the processing of the LLC4320 data (described in \ref{subsec:input output}).}
    \label{tab:features}
\end{table}

\subsection{CNN architecture and training}
\label{sec:NN}

  Each experiment, designed with a given filter scale, is trained and tested independently, but all CNNs shared a common architecture.  We use a CNN architecture for regression inspired by applications for mesoscale eddy parameterizations \cite{ bolton2019applications, guillaumin2021stochastic,perezhogin2023generative}.  A hyper-parameter sweep over the number of hidden layers, kernel size, learning rate, and weight decay, was used to find the best performing CNN. The CNN is trained over 100 epochs while minimizing the Mean Squared Error (MSE) loss (shown in Fig. \ref{fig: loss}). The hyperparameters were tuned against the $1/4^o$ filter scale experiment, and remained fixed throughout all other experiments. Results presented here are based on a CNN with a kernel size of 5X5 in the first layer, followed by 7 hidden convolutional layers with kernel size of 3x3, a learning rate of $2\times10^{-4}$, and weight decay of $0.02$. The total number of learnable parameters is approximately $300,000$. 

 Prior to applying the CNN, all variables listed in Tab. \ref{tab:features} are normalized by a single mean and standard deviation computed over all regions (example shown in Fig. \ref{fig: inputs}). To train the CNN, we randomly select 80\% of the  $\sim 10,000$ samples given from all regions combined. The remaining 20\% is left unseen by the CNN, and is used to test the CNN prediction. Results are compared in the following section with the target LLC4320 data and the \citeA{bodner2023modifying} version of the MLE parameterization, which is used here as a baseline. In subsections \ref{subsec: region} and \ref{subsec: season}, we examine other split choices between the train and test datasets to include subsets of the sampled regions or timeseries, respectively.
 
\section{CNN prediction of subgrid submesoscale fluxes}
\label{sec:results}

Once trained, the CNN has learned a functional mapping, $S(\mathbb{X})$, between the input features, $\mathbb{X}$, and subgrid mixed layer VBF, $\mathbb{Y}_{wb}$. In this section, we examine the extent to which the CNN can make skillful predictions on data that was not included in the training process. For this purpose, we compare the CNN prediction  with the target LLC4320 data held out from training, and test whether the CNN  improves on the baseline given by the \citeA{bodner2023modifying} version of the MLE parameterization. 

\begin{figure}[ht!]
 \centering
\makebox[\textwidth]{\includegraphics[width=1.1\textwidth]
   {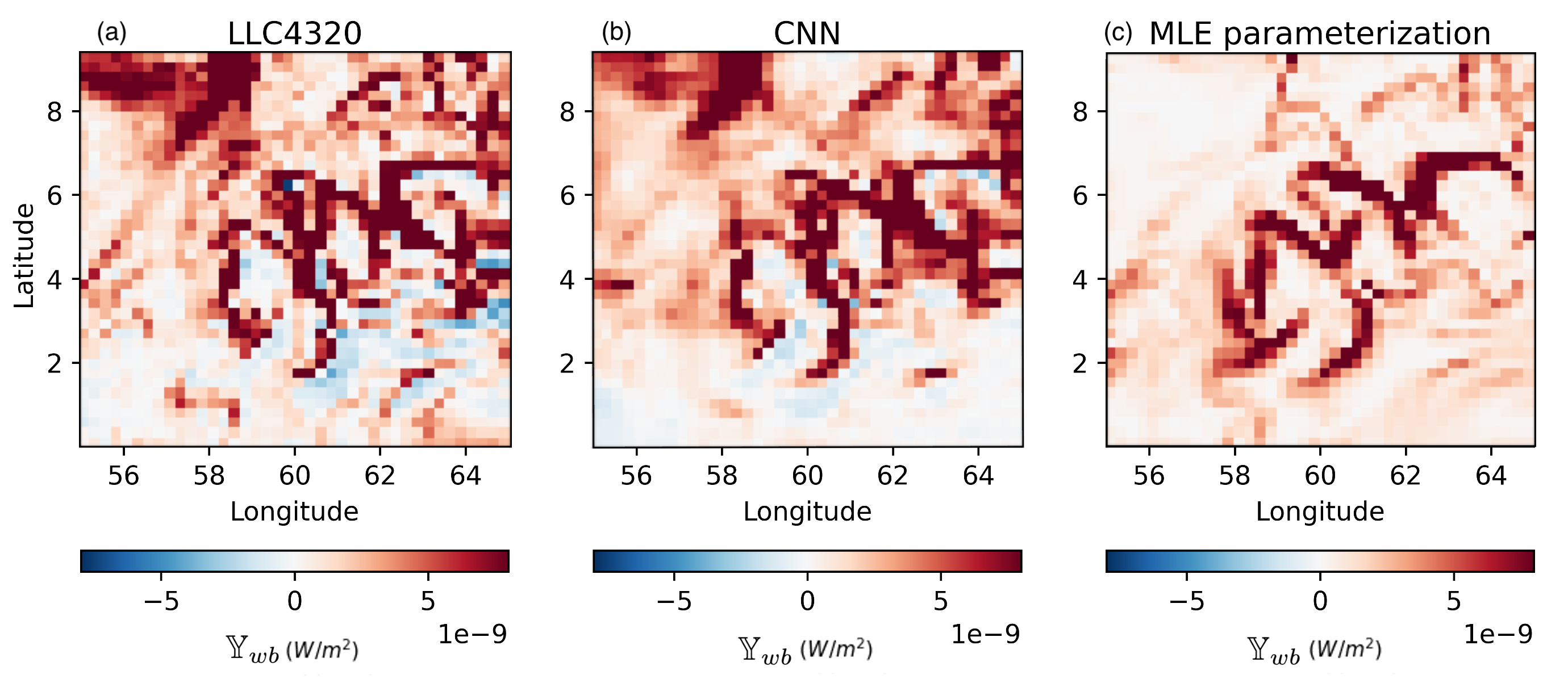}}%
   \\
    \caption{Snapshot taken from the Arabian Sea region of the depth-averaged  subgrid VBF $\mathbb{Y}_{wb}$ [$W/m^2$] given by (a) the LLC4320,  (b) CNN prediction in physical space, and (c) the \citeA{bodner2023modifying} version of the MLE parameterization. A $1/4^o$ filter scale is applied here. An example of (a) with filter scales of $1/12^o, 1/8^o, 1/2^o, 1^o$ is shown in Fig. \ref{fig: snapshot resolution}. }
  \label{fig: predictions snapshot}
\end{figure}

\begin{figure}[ht!]
 \centering
\makebox[\textwidth]{\includegraphics[width=.8\textwidth]
   {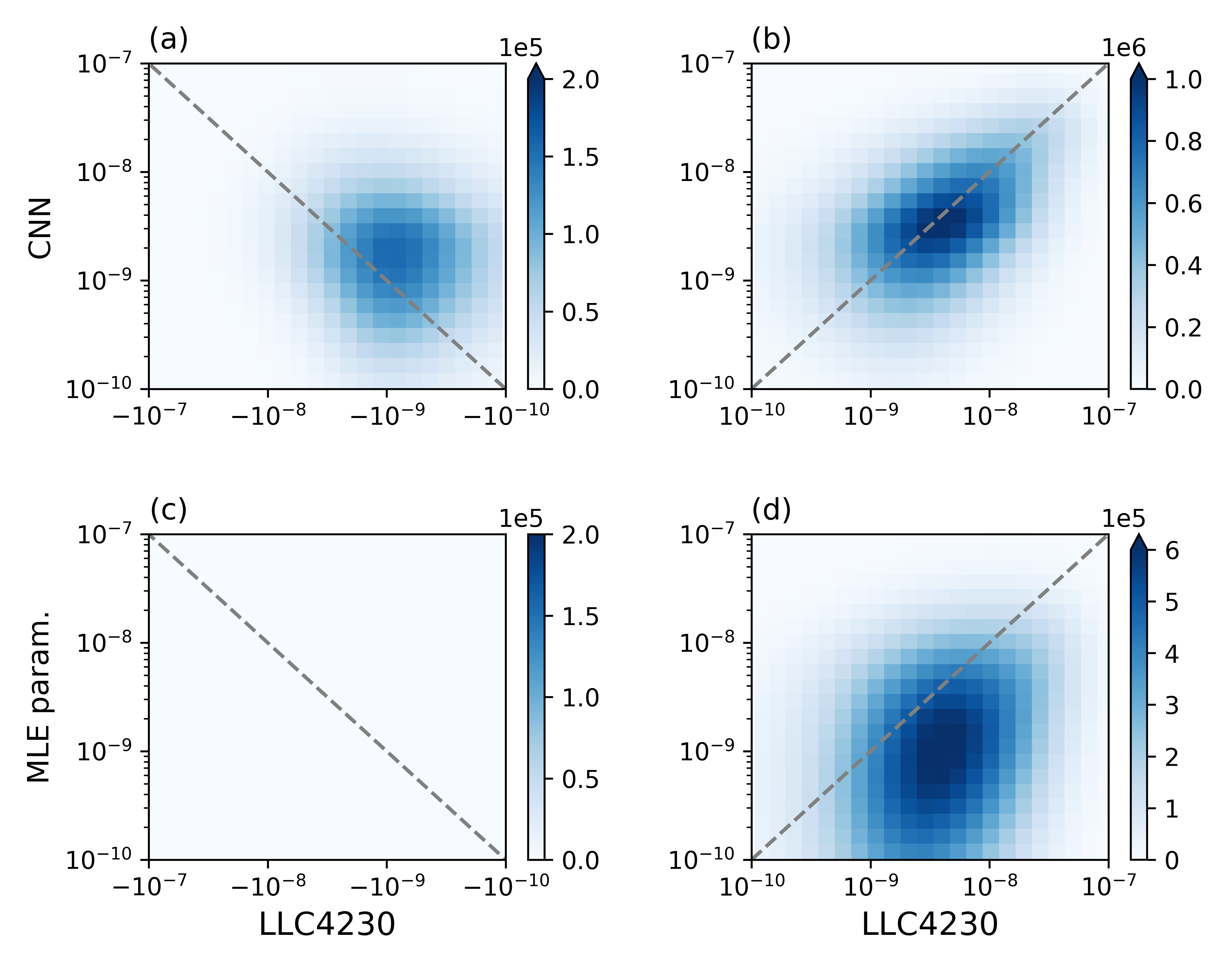}}%
   \\
    \caption{Joint histogram of the $1/4^o$ $\mathbb{Y}_{wb}$ (W/m$^2$): (a,b) CNN prediction and LLC4320 data,  and (c,d) MLE parameterization and LLC4320 data. Panels (a, c) correspond to negative fluxes and (b, d) to positive fluxes. The CNN predictions remain close to the target LLC4320 in both positive and negative values of $\mathbb{Y}_{wb}$. Note that the colorbar in (a) is an order of magnitude larger than the others due to the high concentration along the diagonal.}
  \label{fig: predictions pdf}
\end{figure}

Illustrated by an example from the $1/4^o$ filter scale experiment (Fig. \ref{fig: predictions snapshot}), the CNN is able to capture much of the fine-scale structure and sign of the subgrid VBF. The majority of the fluxes are positive, which is the bulk restratification effect inferred by the MLE parameterization. The negative fluxes exhibited in the LLC4320, and captured by the CNN but not the MLE parameterization, may be indicative of frontogenesis, where an ageostrophic secondary circulation intensifies the front  \cite<e.g.,>[]{hoskins1972atmospheric,shakespeare2013generalized}.  This can be further seen in the joint histogram of the VBF given by the target LLC4320 and those predicted by the CNN and MLE parameterizations (Fig.  \ref{fig: predictions pdf}). The joint histograms are computed over the entire unseen test dataset, which provides a comparison over several orders of magnitude of the VBF. In the case of positive fluxes, the CNN prediction remains close to the LLC4320 VBF, as can be seen by the alignment along the one-to-one gray line, an improvement on the MLE parameterization which deviates from the LLC4320 VBF. For the negative fluxes, the one-to-one alignment is less pronounced, likely due to the significantly smaller number of negative samples seen by the CNN (less than $5\%$ of the total samples). However, the ability of the CNN to predict of negative fluxes is still an improvement on the MLE parameterization, which does not include negative fluxes by construction.  

\begin{figure}[ht!]
   \makebox[\textwidth]{\includegraphics[width=1.2\textwidth]{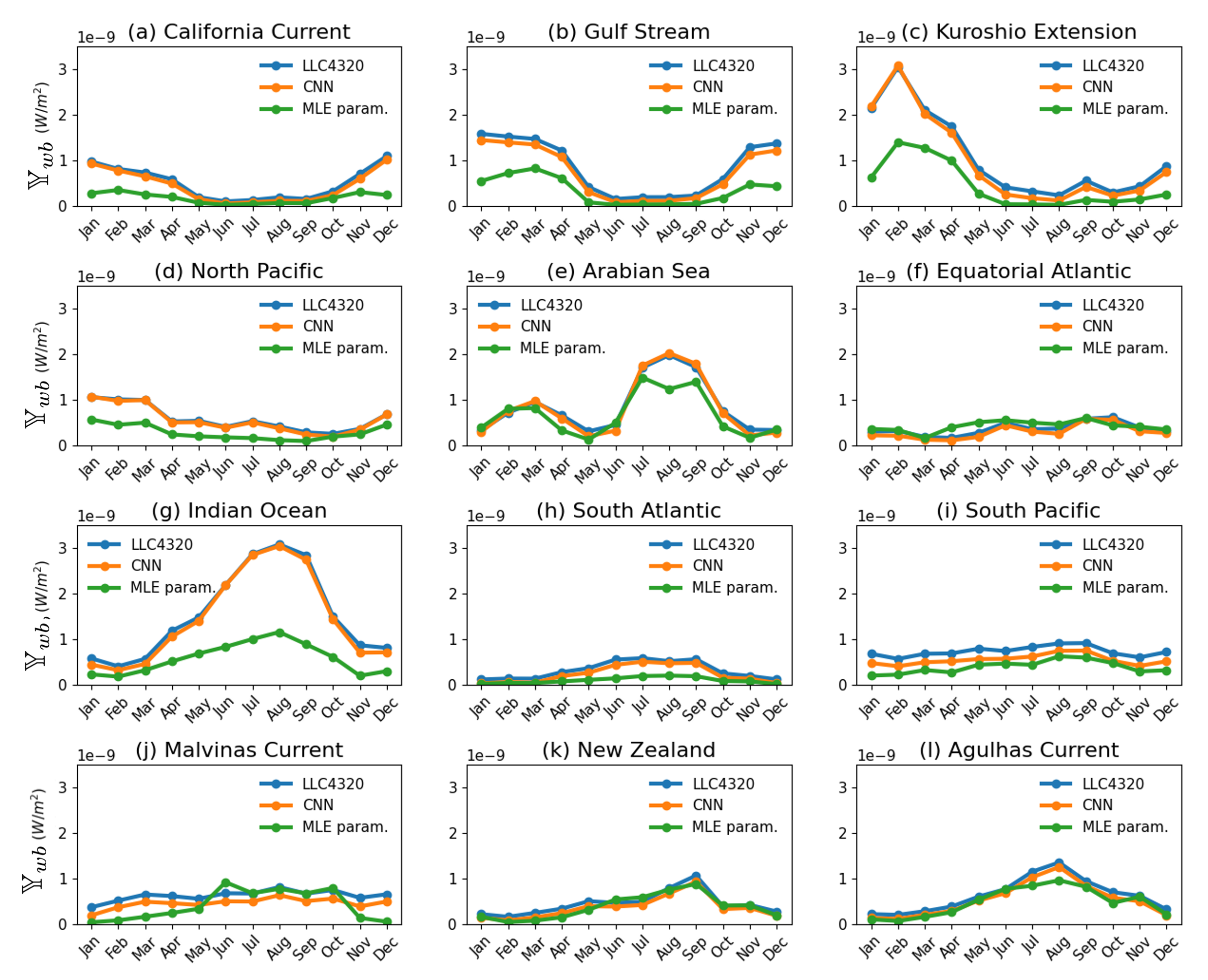}}%
    \caption{Area-weighted spatial average of $\mathbb{Y}_{wb}$ (W/m$^2$) decomposed by region in the $1/4^o$ filter scale experiment. In each panel, CNN predictions of $\mathbb{Y}_{wb}$ on unseen test data (orange) are averaged over each month and compared with the LLC4320 target data (blue) and the MLE parameterization (green). In all regions, the CNN prediction stays close to the LLC4320 target data and surpasses estimates from the MLE parameterization.} 
  \label{fig: prediction timeseries}
\end{figure}

To test whether the CNN also has skill in predicting bulk effects, such as is inferred by the MLE parameterization, the CNN predictions on unseen test data are averaged over each month to form a seasonal cycle, and is compared with the equivalent for the MLE parameterization and LLC4320 target data. We find that in all regions, the CNN prediction captures the seasonality and bulk effects of the LLC4320 data, and outperforms the MLE parameterization, particularly where fluxes appear to be strongest during the winter and spring months where the LLC4320 VBF can be as large as three times the MLE prediction (Fig. \ref{fig: prediction timeseries} for the $1/4^o$ filter scale experiment and more quantitatively in the analysis described below). 

\begin{figure}[ht!]
 \centering
   \includegraphics[width=1\textwidth]{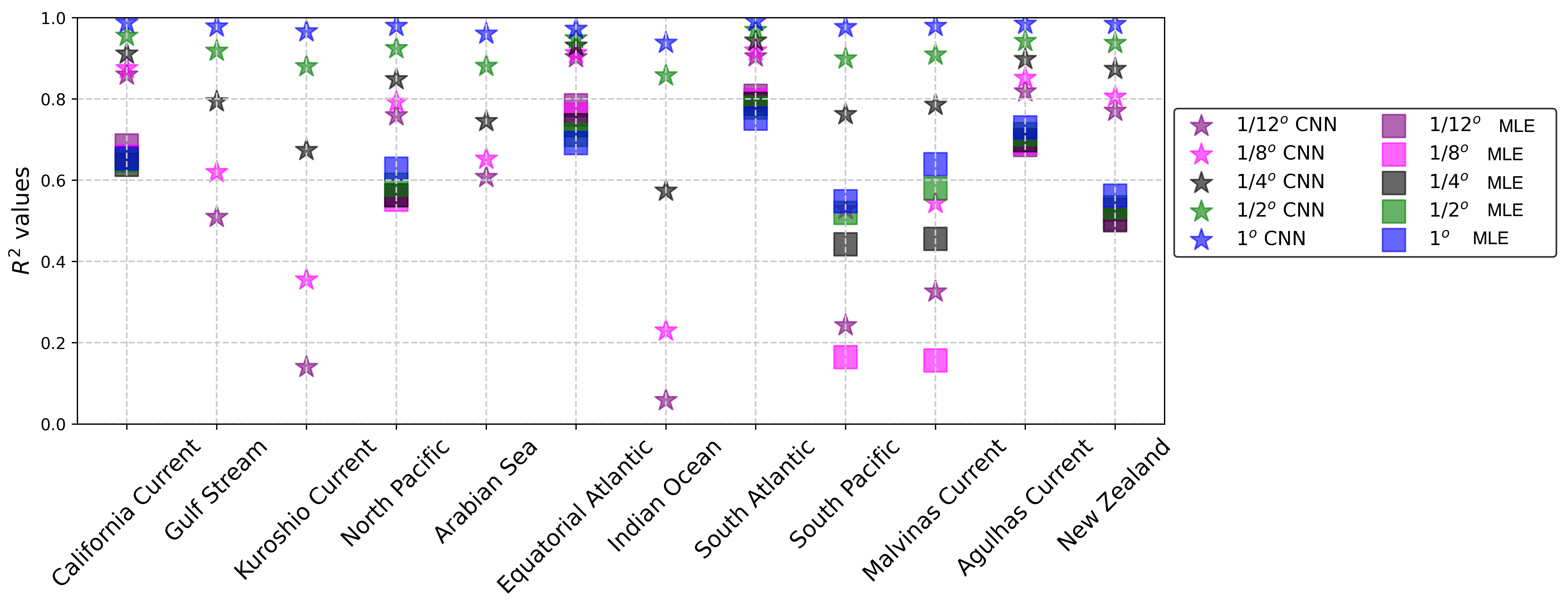}
    \caption{$R^2$ values of CNN prediction on unseen data (stars) and the MLE parameterization estimates (squares) decomposed by regions. Colors represent the different filter scale experiments. Note that $R^2$ is a point-wise estimate and not an averaged quantity as in Fig. \ref{fig: prediction timeseries}. Negative $R^2$ values, exhibited in the South Pacific and Malvinas Current regions in the MLE parameterization, are removed from this figure. The CNN skill exceeds that of the MLE parameterization in all regions and for all filter scales.} 
  \label{fig: r2_resolution_location}
\end{figure}

Prediction skill of the CNN and MLE parameterization for all filter scales are quantified in terms of $R^2$ values relative to the LLC4320 target data (calculation described in more detail in Eq.~\eqref{eq:R2}). We find that the  CNN $R^2$ remain at a value of at least 0.2 higher than that of the MLE in all filter scale experiments and in all regions (Fig.  \ref{fig: r2_resolution_location}). As the magnitude of $\mathbb{Y}_{wb}$ varies spatially  (Fig. \ref{fig: wb cospectrum}), this impacts the predicted output of the CNN in the different regions.  The largest filter scales tend to have skill nearing an $R^2$ values of 1, which then decreases as the filter scale becomes smaller. In the large filter scale experiments, the fields tend to be smoother, as much of the subgrid spatial variability is averaged out, thus presenting an easier learning problem for the CNN. The CNN prediction skill is found to be especially sensitive in the small filter scale experiments, where it performs well in some regions, e.g. in the California Current where all $R^2$ values are above 0.8, but less so in others, e.g. in the Indian Ocean region where the $R^2$ values in the small filter scale experiments are below 0.3).  In the regions that exhibit large sensitivity to the filter scale such as the Kuroshio Current, Indian Ocean, and South Pacific, the skill of the MLE parameterization drops significantly as well. 

It is worth noting that the CNN prediction skill can be sensitive to the initialization weights even when training identical experiment configurations \cite<e.g.,>[]{otness2023data}. However, we find that for each given filter scale experiment, the sensitivity to the initialization weights is smaller than an $R^2$ value of 0.1 (Fig. \ref{fig: init sensitivity}), indicating that the large range of $R^2$ values displayed in Fig. \ref{fig: r2_sensitivity_loc_split} reflects the  sensitivity due to regional variability rather than properties of the CNN.

\begin{figure}[ht!]
 \centering
   \includegraphics[width=1\textwidth]{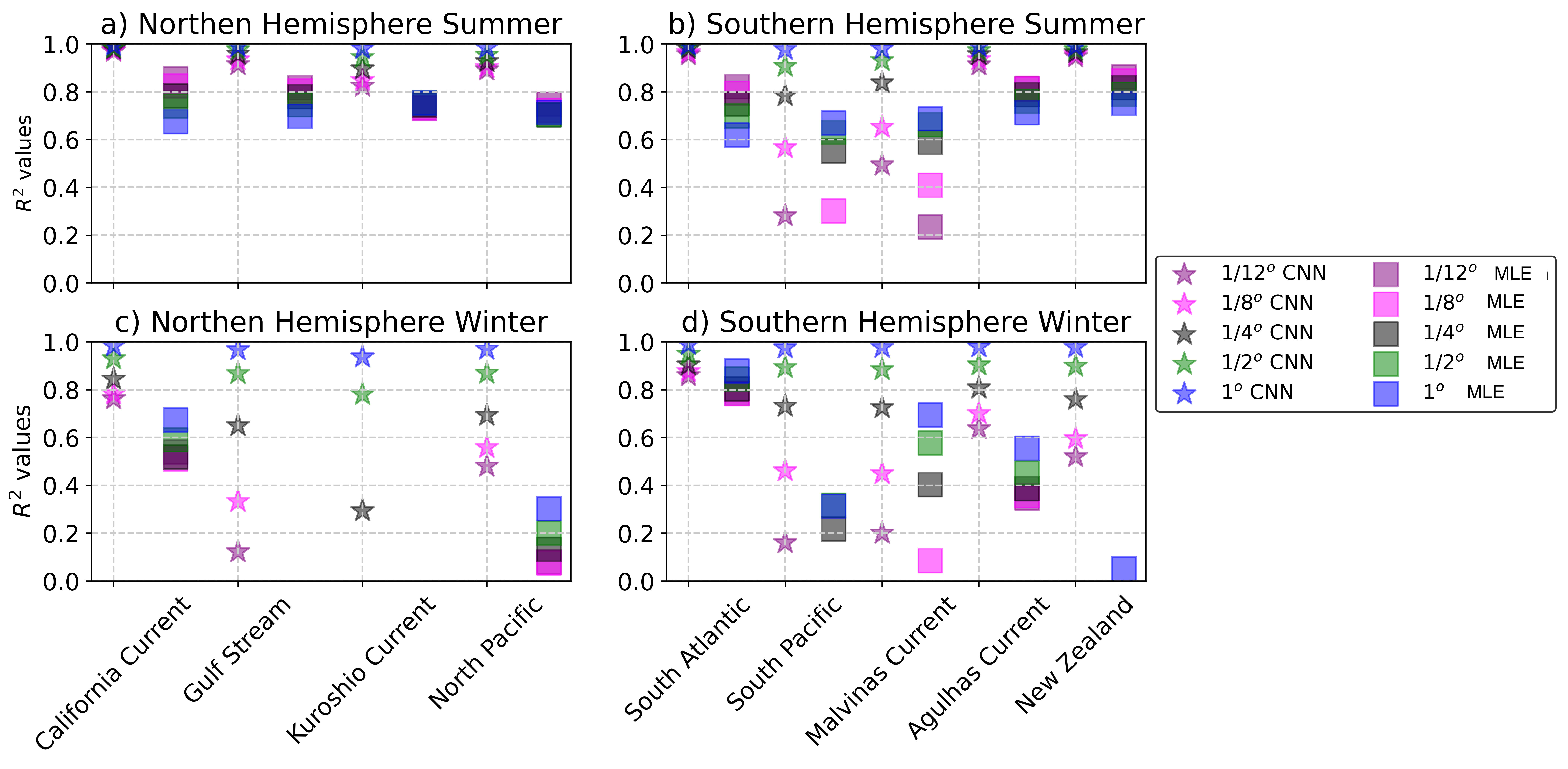}
    \caption{Same as Fig. \ref{fig: r2_resolution_location}, $R^2$ values of CNN prediction on unseen data (stars) and the MLE parameterization estimates (squares) decomposed by regions. Colors represent the different filter scale experiments.  Here we include an average over winter (summer) months: January, February, March and summer (winter) months: July, August, September for regions in the Northern (Southern) Hemisphere. The CNN skill is generally higher in summer (week VBF) compared with winter (strong VBF). Note that we have not included equatorial regions here as the submesoscale equatorial seasonality is less trivial. } 
  \label{fig: r2_cnn_seasons}
\end{figure}

As submesoscale seasonality greatly impacts the variability of VBF (demonstrated in Fig. \ref{fig: prediction timeseries}), we examine the skill (in terms of $R^2$ values) of the CNN and MLE parameterization averaged only over winter and summer months (Fig. \ref{fig: r2_cnn_seasons}). During winter, when mixed layer VBF tend to be stronger, the skill of both the CNN and MLE parameterization in the smaller filter scale experiments drops compared with its equivalent in summer. This is particularly true for regions where fluxes are very strong during winter, such as in the Kuroshio Current, where the $1/8^o$ and $1/12^o$ filter scale experiments shows no skill (negative $R^2$) during winter compared with an $R^2$ value above $0.8$ in summer. Similarly, the Gulf Stream, Agulhas Current, Malvinas Current, and the Southern Ocean near New Zealand all exhibit a drop in skill in both the MLE and CNN with small filter scales. In other regions with less of a pronounced seasonal cycle, such as in the South Pacific or South Atlantic regions, where the CNN skill is roughly the same for summer and winter, and differences appear within the $0.1$ range that can be explained by the CNN initialization sensitivity. Interestingly, the $R^2$ values of the MLE parameterization still drops in both regions during winter in all filter scale experiments. These results suggest that both the CNN and the MLE parameterization struggle to predict the strongest fluxes, generally exhibited during winter and spring months \cite<e.g.,>[]{callies2015seasonality,johnson2016global}. 

To better understand the dependency of our method on the training data, and in particular on regional and seasonal variability, in the following subsections we perform two sensitivity tests by holding out parts of the training data, retraining the CNN, and examining CNN prediction skill on unseen regions or selected fractions of the timeseries. 

\subsection{Holding out regions from training}
\label{subsec: region}
\begin{figure}[ht!]
 \centering
   \includegraphics[width=1\textwidth]{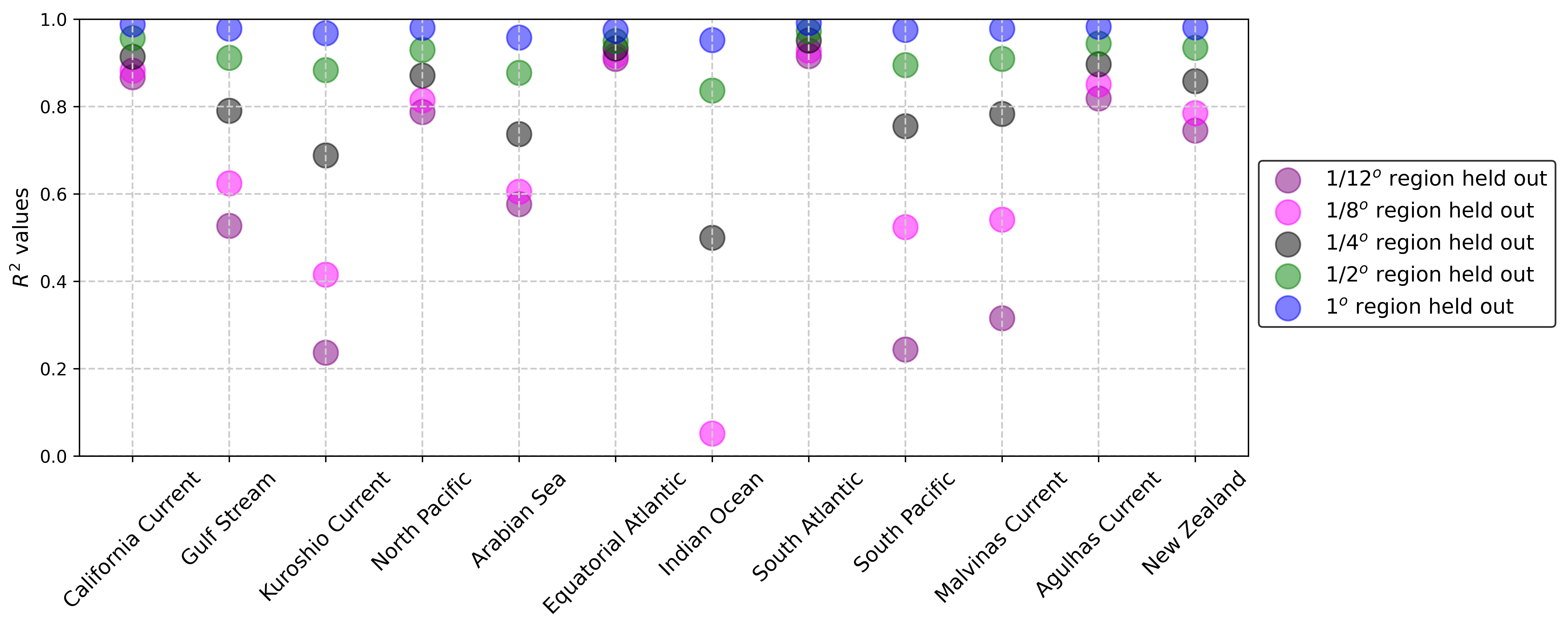}
    \caption{$R^2$ values of CNN prediction on regions held out during training. Colors represent the different filter scale experiments. Results are consistent with Fig. \ref{fig: r2_sensitivity_loc_split} which include all regions for training, suggesting that the CNN is able to generalize onto unseen regions.} 
  \label{fig: r2_sensitivity_loc_split}
\end{figure}

We test the ability of the CNN to make predictions on regions that are not included in the training data. We thus generate 12 new datasets that correspond to removing one region at a time from the training dataset. We retrain the CNN in 12 different experiments, and make predictions on a different unseen region each time.   The $R^2$ values of the CNN on the unseen regional data (Fig. \ref{fig: r2_sensitivity_loc_split})  remain consistent with those found on the full training set (Fig. \ref{fig: r2_resolution_location}) across filter scales and over all regions. This suggests that the training data covers a wide enough range of dynamical regimes that enables generalization of the CNN  on regions not included in training, an especially important result given that a fairly small number of regions were included in training compared with the full ocean.

\subsection{Holding out seasonality from training}
\label{subsec: season}

 We perform two experiments in which we hold out winter and summer months from the training data, to examine the ability of the CNN to make predictions on unseen seasonal variability. We thus create two new training and test datasets to better understand the overall sensitivity of our method to submesoscale seasonality: \begin{itemize}
    \item \textbf{Winter held out} refers to training data which excludes from the time series the months of January, February, March from all regions in the Northern Hemisphere, and July, August, September from regions in the Southern Hemisphere. Note again that we have removed equatorial regions from the analysis entirely. The remainder of the time series-- e.g. spring, summer, fall-- is used to train the CNN, and predictions are made on the unseen winter data. 
    \item \textbf{Summer held out} is same as the above, where we now exclude July, August, September  from the Northern Hemisphere and January, February, March from the Southern Hemisphere. Equatorial regions are once again excluded.
\end{itemize}

\begin{figure}[ht!]
 \centering
   \includegraphics[width=1\textwidth]{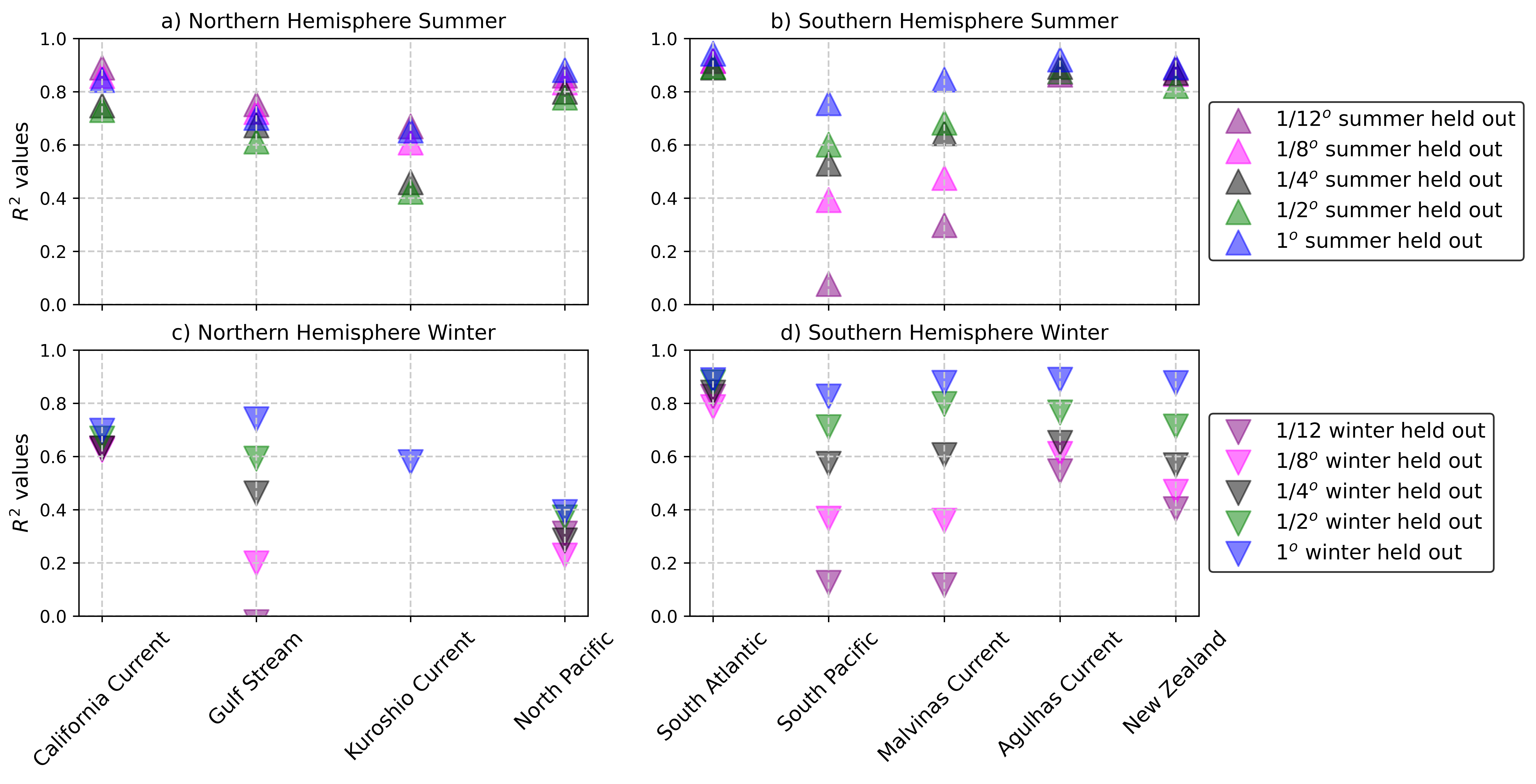}
    \caption{ $R^2$ values of CNN prediction on seasons held out during training: summer (upward triangle), winter (downward triangle). Colors represent the different filter scale experiments.  Same as Fig. \ref{fig: r2_cnn_seasons}, we average over winter (summer) months: January, February, March and summer (winter) months: July, August, September for regions in the Northern (Southern) Hemisphere.  The CNN skill in the Northern Hemisphere regions, in particular near strong ocean boundary current, drops compared with its equivalent in Fig. \ref{fig: r2_sensitivity_seasons_split}. } 
  \label{fig: r2_sensitivity_seasons_split}
\end{figure}

In these experiments, we find that results differ between regions in the Northern and Southern Hemispheres. In the Northern Hemisphere regions, the $R^2$ values in the "Summer held out" experiments decrease by a margin larger than 0.1 compared with the CNN trained on the full timeseries (Fig. \ref{fig: r2_sensitivity_seasons_split}a compared with Fig. \ref{fig: r2_cnn_seasons}). We find the largest decreases in particular in the small filter scale experiments in regions affected by strong ocean boundary currents (i.e., the Gulf Stream and Kuroshio Current). Contrarily, the Southern Hemisphere regions are found to be consistent with the  predictions for experiments trained on the full timeseries (Fig. \ref{fig: r2_sensitivity_seasons_split}b and Fig. \ref{fig: r2_cnn_seasons}). The CNN predictions in the "Winter held out" experiments result in lower  $R^2$ values in the Northern Hemisphere (Fig. \ref{fig: r2_sensitivity_seasons_split}c) compared with predictions trained on the the full timeseries. In the Kuroshio Current for example, all filter scales smaller than $1^o$ result in negative $R^2$ values, indicating that there is no skill in the CNN prediction in these cases. In the Southern Hemisphere regions (Fig. \ref{fig: r2_sensitivity_seasons_split}d), the "Winter held out" experiments also display a decrease in $R^2$ values but it is still within the margin that can be explained by sensitivity to the initialization weights (Fig. \ref{fig: init sensitivity}).  These results reinforce the need for the distribution of the data used to train the CNN to include the strongest seasonal fluxes, and particularly from regions where submesoscales are most active, such as near ocean boundary currents.

Results from both the seasonal and regional sensitivity experiments indicate that the learned relationships between the input features and $\mathbb{Y}_{wb}$ can extend over a variety of dynamical regimes, especially in the large filter scale experiments. In the following section we delve deeper in attempt to interpret these relationships.

\section{Local and non-local feature importance}
\label{sec:features}
We have shown that the CNN improves on the MLE parameterization, but an important remaining question is why? What relationships are learned between the input variables and $\mathbb{Y}_{wb}$ that lead to better predictions by the CNN? With such complex and nonlinear relationships, it is difficult to decipher which input feature is most important and for what reason. Many methods exist that help explain and interpret the dependency of CNN outputs to its inputs \cite<e.g.,>[]{ zeiler2014visualizing, ribeiro2016should,  selvaraju2017grad, van2022tractability}. Here, we have chosen two complimentary methods that help gain insight on the learned relationships and the importance of individual inputs to $\mathbb{Y}_{wb}$.

\subsection{Impact of input feature on CNN prediction skill}
\begin{figure}[ht!]
 \centering
   \includegraphics[width=\textwidth]{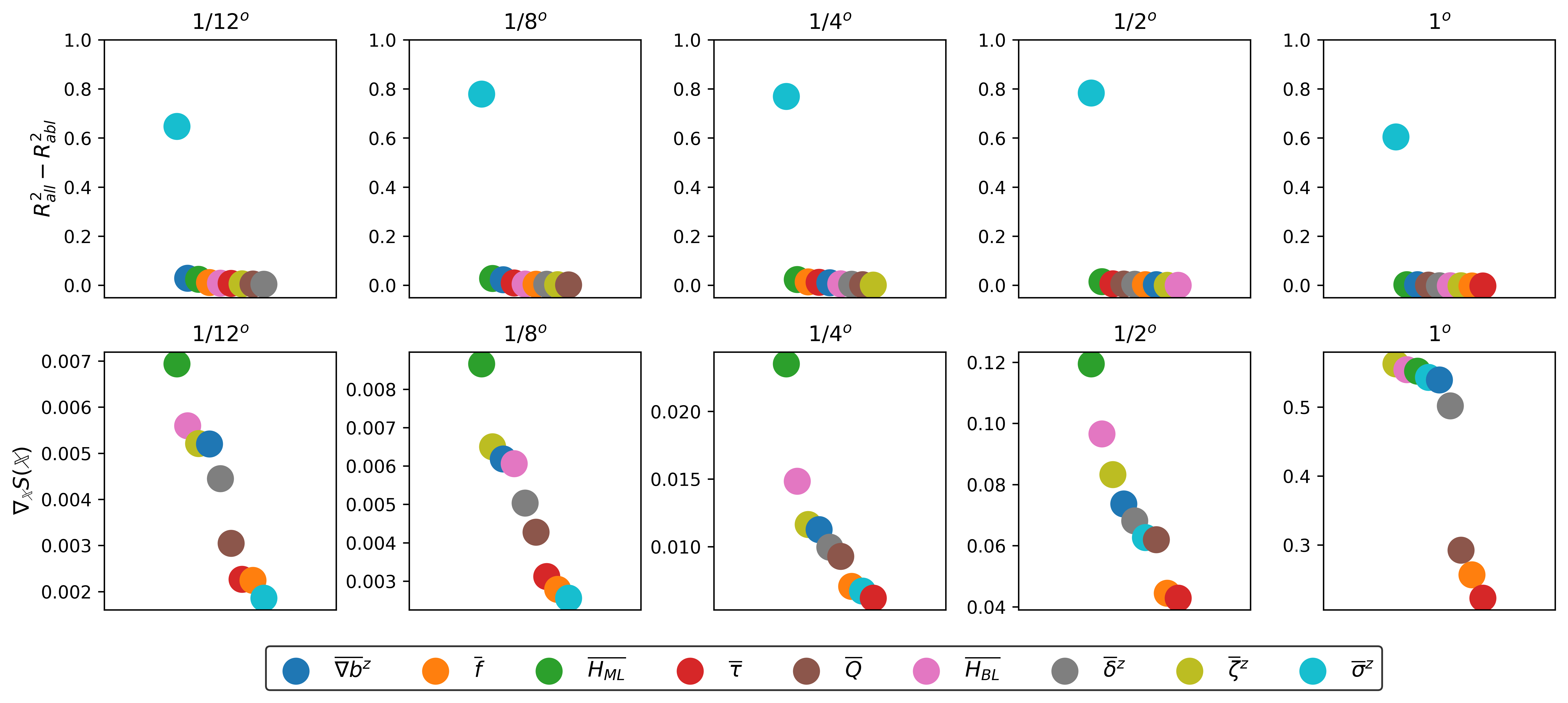}
    \caption{Explainability methods: Top panels are the relative $R^2$ values, ${R_{all}^2-R_{abl}^2}$, between the CNN containing all input features and results from the ablation experiment, where one input feature is removed at a time.  Bottom panels are the Jacobian $\nabla_{\mathbb{X}}S(\mathbb{X})$ of the output with respect to individual inputs (in normalized units). In both methods, a high score indicates sensitivity to input features. Columns represent the filter scale experiments. Note that the entire unseen dataset (including all regions and seasons) was used here. } 
  \label{fig: r2_feature_jacobian}
\end{figure}

To test the dependency of the CNN prediction on certain input features, we perform a set of ablation experiments, where we remove one input feature at a time, retrain the CNN, and examine the resulting prediction skill in terms of the relative $R^2$ value. This relative $R^2$ value is taken as the difference between $R_{all}^2$, resulting from the experiment with all input features included, and $R_{abl}^2$, resulting from the ablation experiment for each input. A  high $R_{all}^2-R_{abl}^2$ value indicates that the skill has dropped in a particular ablation experiment, meaning that the CNN strongly depends on said input feature (top panels in Fig. \ref{fig: r2_feature_jacobian}). Notably, strain demonstrates the strongest dependency of the CNN consistently across all filter scales. Interestingly, there appears to be very little sensitivity to the removal of any other input feature, including those used by the MLE parameterization.

\subsection{Sensitivity of output  relative to input features}
\label{subsec: jacobian}
We next apply a complimentary method to the ablation experiment above. The Jacobian of the CNN prediction, $S(\mathbb{X}$), is computed with respect to the input features by taking gradients along the CNN weights, $\nabla_{\mathbb{X}}S(\mathbb{X})$. The Jacobian is an especially useful metric to evaluate the point-wise sensitivity of the output to each input feature \cite<e.g.,>[]{ross2023benchmarking}. Note that unlike the ablation experiment, where we examined the  $R^2$ value on the full output domain, the Jacobian considers only the sensitivity of a single output grid cell to a single grid cell in the input feature map.  Here, we compute the Jacobian over the entire unseen test dataset, and examine its \textit{average} values for each input feature, thus providing a metric for how sensitive, on average, the CNN output is to each input feature. We contrast the Jacobian with the $R_{all}^2-R_{abl}^2$ values given by the ablation experiments (Fig. \ref{fig: r2_feature_jacobian}), where for the Jacobian, a high score indicates that the CNN prediction, $S(\mathbb{X})$, is  sensitive to point-wise changes in a certain input. We find that the highest-ranked input feature, for which $S(\mathbb{X})$ is most sensitive to, is the mixed layer depth, $H_{ML}$, which is generally a one-dimensional, local property determined by surface forcing. The sensitivity to mixed layer depth is followed by sensitivity to boundary layer depth, the buoyancy gradient, and vorticity. Note that $S(\mathbb{X})$ does not appear to be sensitive to point-wise changes in surface heat flux, surface wind stress, or Coriolis, which is likely due to these fields being smoother in the LLC4320 at the scales relevant for the Jacobian. Despite strain being the most important feature in the previous section, it is only in the $1^o$ filter scale experiment that the Jacobian exhibits sensitivity of $S(\mathbb{X})$ to vorticity, divergence, and strain, indicating that the impact of these fields is most apparent at the large scale. 

\subsection{Receptive field of the CNN}
\begin{figure}[ht!]
 \centering
\makebox[\textwidth]{\includegraphics[width=1.1\textwidth]{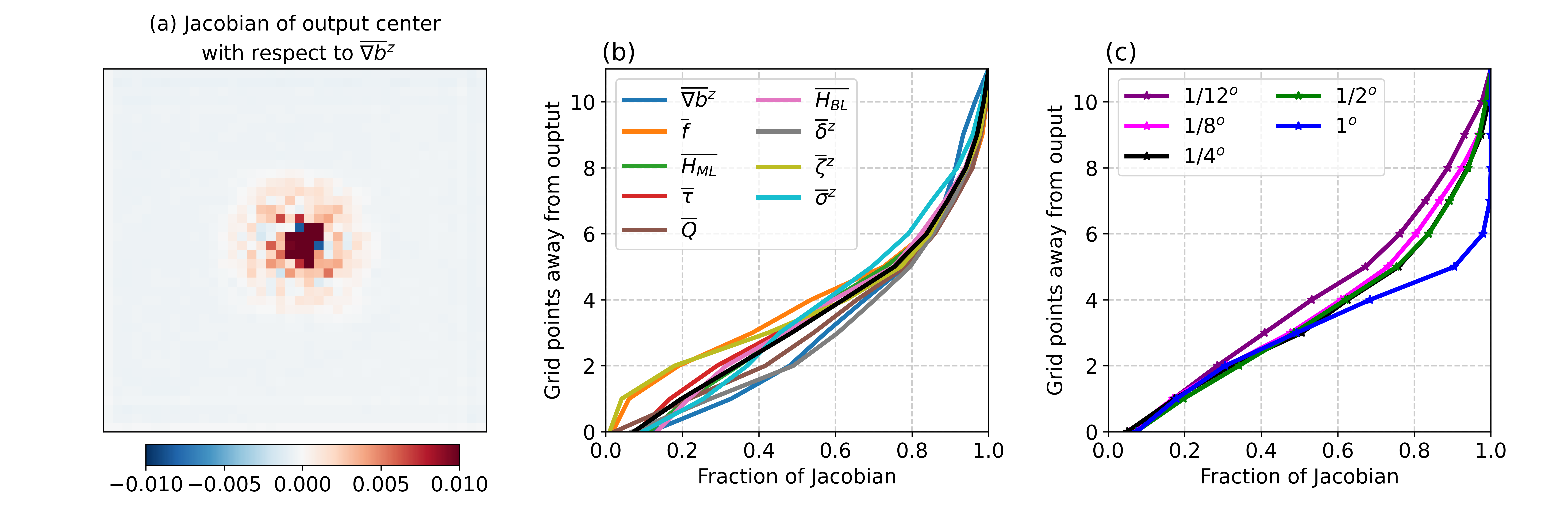}}%
    \caption{(a) Example of the Jacobian at the center of the input domain with respect to the  buoyancy gradient field in a $1/4^o$ filter scale experiment. (b) Radial grid point number compared with fraction of $\nabla_{\mathbb{X}}S(\mathbb{X})$ for all input features in the $1/4^o$ filter scale experiment. Black line is the mean over all inputs. (c) Average fraction of Jacobian for all filter scales experiments. We find that 7 grid points captures $90\%$ of the Jacobian fraction, which corresponds to the number of grid points required to capture sensitivity  between the input map and a single output grid cell. In all panels the Jacobian is computed in normalized units.} 
  \label{fig: jacobian_feature_importance_halo}
\end{figure}

To further understand the relevance of locality, we follow the analysis in \citeA{ross2023benchmarking} for examining the Jacobian of the output center point with respect to the full domain of each input feature. An example for the buoyancy gradient input feature is shown in Fig. \ref{fig: jacobian_feature_importance_halo}a, where the shaded area illustrates the CNN's receptive field needed to predict a single output point. Averaging over that halo, we examine the fraction of Jacobian over the number of grid points, which can be thought of as the percentage of sensitivity for each input feature that is being captured by the CNN (Fig. \ref{fig: jacobian_feature_importance_halo}b,c). We find that 7 grid points away from the center is sufficient for capturing 90\% of the Jacobian fraction, e.g.  90\% of sensitivity between the output and input features. A relatively local receptive field is found to be consistent across filter scales despite the varying importance of input features found previously. 

This result complements the analysis in \citeA{gultekin2024analysis}, which found that the \citeA{guillaumin2021stochastic} CNN skill saturates at a stencil of seven grid points for coarse-graining factors of 4, 8, 12, 16. As discussed in \ref{sec:NN}, the CNN architecture choice used here is motivated by that used in \citeA{guillaumin2021stochastic}, however, the physical phenomena we are parameterizing is different, and so are the input and output features. An investigation of the significance of a seven grid point stencil and its dependence on the CNN architecture in both cases is left for a future study.

\section{Discussion and Conclusions}
\label{sec:conclusion}

The parameterization for submesoscale VBF plays a key role in setting stratification in the ocean mixed layer, and as such contributes to the exchange between the ocean and atmosphere systems in GCMs. In contrast to previous physics-based approaches, here we develop a new data-driven parameterization, where a CNN is trained to predict mixed layer VBF. The subgrid flux, $\mathbb{Y}_{wb}$, is inferred by the CNN as a function of 9 large-scale input features with known relevance to submesoscale VBF: $\overline{\nabla b}^z$, $f$, $H_{ML}$, $N^2$, $Q$, $\tau$, $H_{BL}$, $\overline{\sigma}^z$, $\overline{\delta}^z$, $\overline{\zeta}^z$ (see Tab. \ref{tab:features}). The data used for training is given from 12 regions sampled from the global high-resolution LLC4320 simulation output. The CNN is trained over a random selection of $80\%$ of all data, while the remaining $20\%$ is unseen by the CNN and is used for testing. We perform five filter scale experiments of $1/12^o,1/8^o,1/4^o,1/2^o,1^o$ and compare with a baseline given by the \citeA{bodner2023modifying} formulation of the MLE parameterization. 

We consistently find that the CNN predictions improve on the MLE parameterization, with higher $R^2$ values across all regions, seasons, and filter scales tested in this study. We additionally perform several sensitivity experiments, where we test the CNN's ability to make predictions on regional or temporal data held out during training. It is found that the CNN, in particular in the larger filter scale experiments, is able to make skillful predictions on unseen data, but does poorly during when submesoscales are most active,  generally corresponding to winter months and near ocean boundary currents. 

The significant improvement on the MLE parameterization indicates that the CNN has learned new meaningful relationships between the input features and $\mathbb{Y}_{wb}$, such that it is able to make skillful predictions over widely different dynamical regimes. We applied two complimentary explainability methods which enable a closer look at the relationships between the CNN output and input features. We find that the CNN exhibits strong dependency on the local relationship between $\mathbb{Y}_{wb}$ and the mixed layer depth, a 1D property driven by surface forcing, and strong non-local dependency on the large scale strain field, a variable closely related to frontogenesis, which is currently missing from the MLE parameterization in GCMs. 

An important limitation of these method is in detecting the importance of  features that are strongly correlated with other fields, such that removing one feature may not lead to differences in the CNN prediction skill. In particular, fields associated with surface forcing, e.g., mixed layer, boundary layer, surface heat flux, and wind stress are expected to be correlated (as is also found to be true in the LLC4320 data, Fig. \ref{fig: corr map}). Nonetheless, results from the ablation experiments suggest that the primary reason the CNN predictions surpass those of the MLE parameterization are due to the newly-captured non-local relationship between $\mathbb{Y}_{wb}$ and the large scale strain field, on which the CNN is strongly dependent. Note that strain is known for its role in constraining submesoscale fluxes and contributing to frontal intensification \cite<e.g.,>[]{shcherbina2013statistics, balwada2021vertical,sinha2023submesoscales}, but these findings emphasize the relevance of strain to improving submesoscale VBF parameterizations, such as recently proposed by \citeA{zhang2023parameterizing}. An extension of these approaches may include incorporating more data from other submesoscale permitting simulations \cite{uchida2022cloud,gultekin2024analysis}, or an investigation of causal links which are not captures in the current framework \cite<e.g.,>[]{camps2023discovering}. An equation discovery approach \cite<e.g.,>[]{zanna2020data} may enable a closer comparison with \citeA{zhang2023parameterizing}, and whether the relationship between strain and submesoscale fluxes emerge in a similar fashion.

We have demonstrated that the CNN improves on the MLE parameterization in an offline setting. However, it is important to note that during training, the CNN minimizes the MSE loss-- a metric closely related to the $R^2$ value (as shown in Eq.~\eqref{eq:R2}). The MLE parameterization, on the other hand, is designed to represent the bulk effects of submesoscale VBF. A next important step is to explore the implications of better captured mixed layer VBF in a GCM and compare with the MLE parameterization online, such as in a recent attempt by \citeA{zhou2024neural} in Regional Ocean Modeling System (ROMS). We have designed our method to correspond with the existing implementation of the MLE parameterization in GCMs, where the theoretical expression for $\mathbb{Y}_{wb}$ in Eq.~\eqref{eq:wb_FK} can simply be replaced with the CNN. A relatively small receptive field of 7 grid points is found to be sufficient at capturing relationships between the input features and $\mathbb{Y}_{wb}$, which suggests that a smaller network may aid future implementation efforts in GCMs \cite{zhang2023implementation}. A decomposition may be preferred to distinguish the bulk restratification effect with the intermittent negative fluxes, and will allow a more natural relationship with vertical buoyancy fluxes already estimated in boundary layer turbulence parameterizations \cite{large1994oceanic,reichl2018simplified, sane2023parameterizing}. The exact formulation, implementation, and evaluation of impact on climate variables is left for future work. 

Beyond the modeling framework discussed above, the utility of our work can also be made amenable to observational data. In particular, the Surface Water and Ocean Topography (SWOT) altimeter mission is starting to provide measurements of sea surface height at an unprecedented resolution \cite{morrow2019global}. The data-driven approach presented here provides an opportunity to leverage surface fields derived from SWOT \cite<e.g.,>[]{qiu2016reconstructability,bolton2019applications}, to infer subsurface VBF and gain new insights of upper ocean dynamics.


\section*{Open data}
Data from the LLC4320 simulation can be accessed using the {\tt llcreader} Python package \cite{abernathey2019petabytes}. Variables from the LLC4320 output in the regions used in this study are stored on the LEAP-Catalog \url{https://catalog.leap.columbia.edu/feedstock/highresolution-ocean-simulation-llc4320-12hourly-averaged-3d-regions}. Code used to process the LLC4320, train the CNN, and generate the figures in this manuscript can be found at \url{https://github.com/abodner/submeso_param_net}. Diagnostics incorporate open source Python packages: { \tt xhistogram}, \url{xhistogram.readthedocs.io}; {\tt fastjmd95}, \cite{abernathey2020fastjmd95} ; {\tt xmitgcm},  \cite{abernathey2021xgcm}.

\section*{Acknowledgements}
 This research received support through Schmidt Sciences, LLC. AB was supported by a grant from the Simons Foundation: award number 855143, Bodner. We thank members of the M$^2$LInES project for support and constructive feedback during the formulation of ideas, in particular, Pavel Perezhogin, Chris Pedersen, Ryan Abernathey, Carlos Fernandez-Granda, and Fabrizio Falasca. The authors would also like to thank the Pangeo Project for providing open-source code which enabled timely analysis for working with the LLC4320 data. This research was also supported in part through the NYU IT High Performance Computing resources, services, and staff expertise.
\newpage

\bibliographystyle{agusample}
\bibliography{main}

\renewcommand{\thefigure}{S\arabic{figure}}
\renewcommand{\theequation}{S\arabic{equation}}
\renewcommand{\thesection}{S\arabic{section}}
\renewcommand{\thetable}{S\arabic{table}}
\setcounter{figure}{0}
\setcounter{section}{0}
\setcounter{table}{0}
\setcounter{equation}{0}

\section{Calculation of $R^2$ values}
The $R^2$ metric is computed by,
\begin{equation}
    R^2= 1-\frac{\sum_i \left(\mathbb{Y}_{wb,i} - \hat{\mathbb{Y}}_{wb,i} \right)^2}{\sum_i\left(\mathbb{Y}_{wb,i} - <\mathbb{Y}_{wb}> \right)^2}\ ,
    \label{eq:R2}
\end{equation}

\noindent where $\mathbb{Y}_{wb,i}$ is the target subgrid flux (at location $i$) as defined in Table \ref{tab:features}, $\hat{\mathbb{Y}}_{wb,i}$ is the same for prediction of $\mathbb{Y}_{wb}$ by the CNN or MLE parameterization, and\\ $<\mathbb{Y}_{wb}> = \frac{1}{n}\sum_{i=1}^{n}$ $\mathbb{Y}_{wb,i}$ is the mean of $\mathbb{Y}_{wb}$ over all $n$ samples. 

\section{The MLE parameterization}

The MLE parameterization is cast in the form of a streamfunction $\boldsymbol{\Psi}_{MLE}$ \cite{fox2008parameterization}, provided by a scaling for,
\begin{equation}
\overline{w'b'}^{z}\propto \frac{H_{ML}|\overline{\nabla_H b|}^z}{|f|},
\label{eq:wb_FK}    
\end{equation}

 \noindent where $H_{ML}$ is the mixed layer depth,  $f$ is the Coriolis parameter, $w$ is vertical velocity, $b$ is buoyancy, and $\nabla_H b$ is the horizontal buoyancy gradient. We follow the notation in \citeA{fox2008parameterization}, where the horizontal spatial resolution of the GCM is denoted $\overline{(\cdot)}$ and $(\cdot)'$ is the unresolved subgrid variable. Superscript $z$ represents a vertical average over the mixed layer depth. The scaling for submesoscale vertical buoyancy flux, $\overline{w'b'}^{z}$, is derived from the bulk extraction of potential energy by mixed layer eddies. A shape function $\mu(z)$ estimates the depth, $z$, at which the mixed layer eddy fluxes are activated,
    \begin{equation}
        \mu(z) = \max\left(0, \left[1-\left( \frac{2z}{H_{ML}}+1\right)^2 \right] \left[1+\frac{5}{21}\left( \frac{2z}{H_{ML}}+1\right)^2 \right]\right),
        \label{eq:structure function}
    \end{equation}
where $\mu(z)$ is set to vanish at the surface and below the mixed layer $H_{ML}$.

\begin{table}[h]
    \centering
\begin{tabular}{l l l l}
\textbf{Region}  & \textbf{Latitudinal Range} & \textbf{Longitudinal Range} \\
1. California Current	& 	(30,45) &	(-140,-125)\\
2. Gulf Stream &	(30,45) & 	(-60,-45) \\
3. Kuroshio Extension &	(25,40) & (145,160) 	\\
4. North Pacific	&	(10,25)	& (-180,-165)\\
5. Arabian Sea	&	(0,15) &	(55,70) \\
6. Equatorial Atlantic	&	(-8,8)	& (-30,-15)\\
7. Indian Ocean	&	(-25,-10) &	(70,85)\\
8. South Atlantic	&	(-30,-15) &	(-25,-10)\\
9. South Pacific	&	(-45,-30) &	(-140,-125)\\
10. Malvinas Current	&	(-55,-40) &	(-60,-45)\\
11. Agulhas Current	&	(-55,-40) &	(20,35)\\
12. Southern Ocean, New Zealand	&	(-60,-45) &	(-175,-160)\\

\bigskip

\end{tabular}
\caption{Coordinate range of sampled regions from the LLC4320 used in this study, corresponding to the blue boxes in Figure \ref{fig: grad b}.}
\label{tab:regions}
\end{table}

\begin{figure}[ht!]
 \centering
   \includegraphics[width=.6\textwidth]{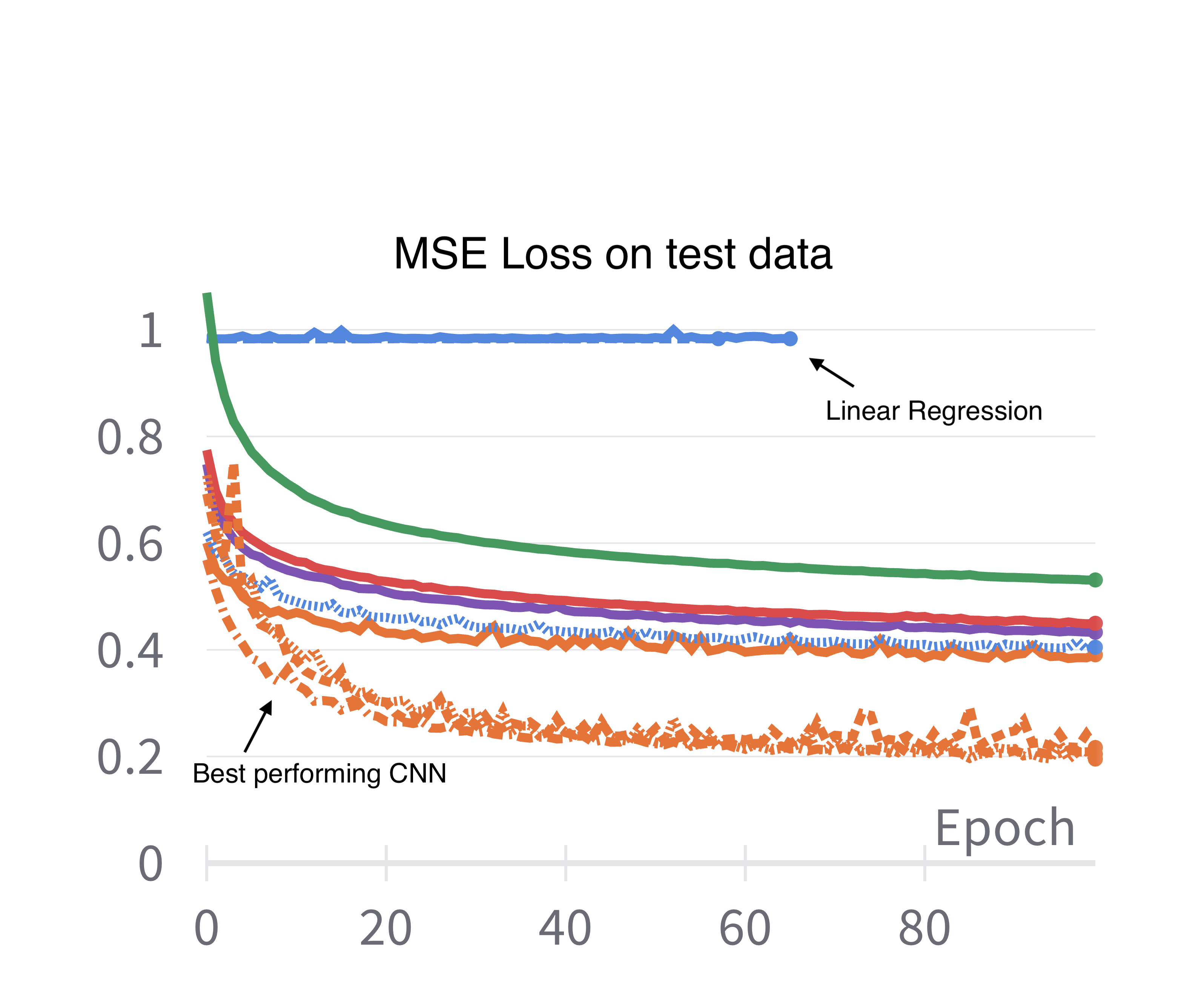}
    \caption{Example of the loss function during the hyper-parameter sweep of the CNN in the $1/4^o$ filter scale experiment. Solid blue line is the case of a simple linear regression, which is not sufficient to reduce the MSE loss.} 
  \label{fig: loss}
\end{figure}

\begin{figure}[ht!]
 \centering
   \makebox[\textwidth]{\includegraphics[width=1.1\textwidth]{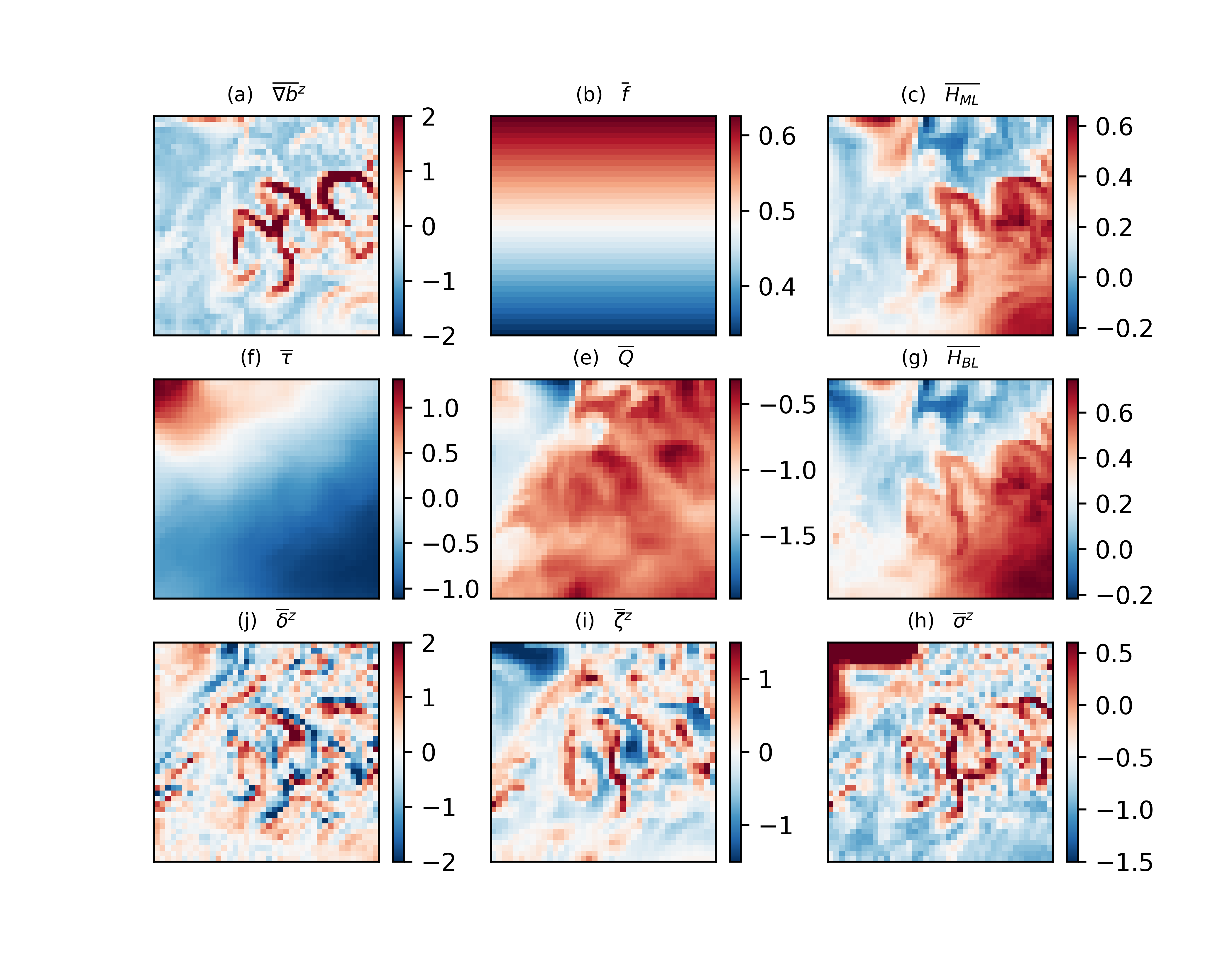}}%
    \caption{Snapshots of the normalized input features (corresponding to Table \ref{tab:features}) in the  $1/4^o$ filter scale.} 
  \label{fig: inputs}
\end{figure}

\begin{figure}[ht!]
 \centering
   \includegraphics[width=.8\textwidth]{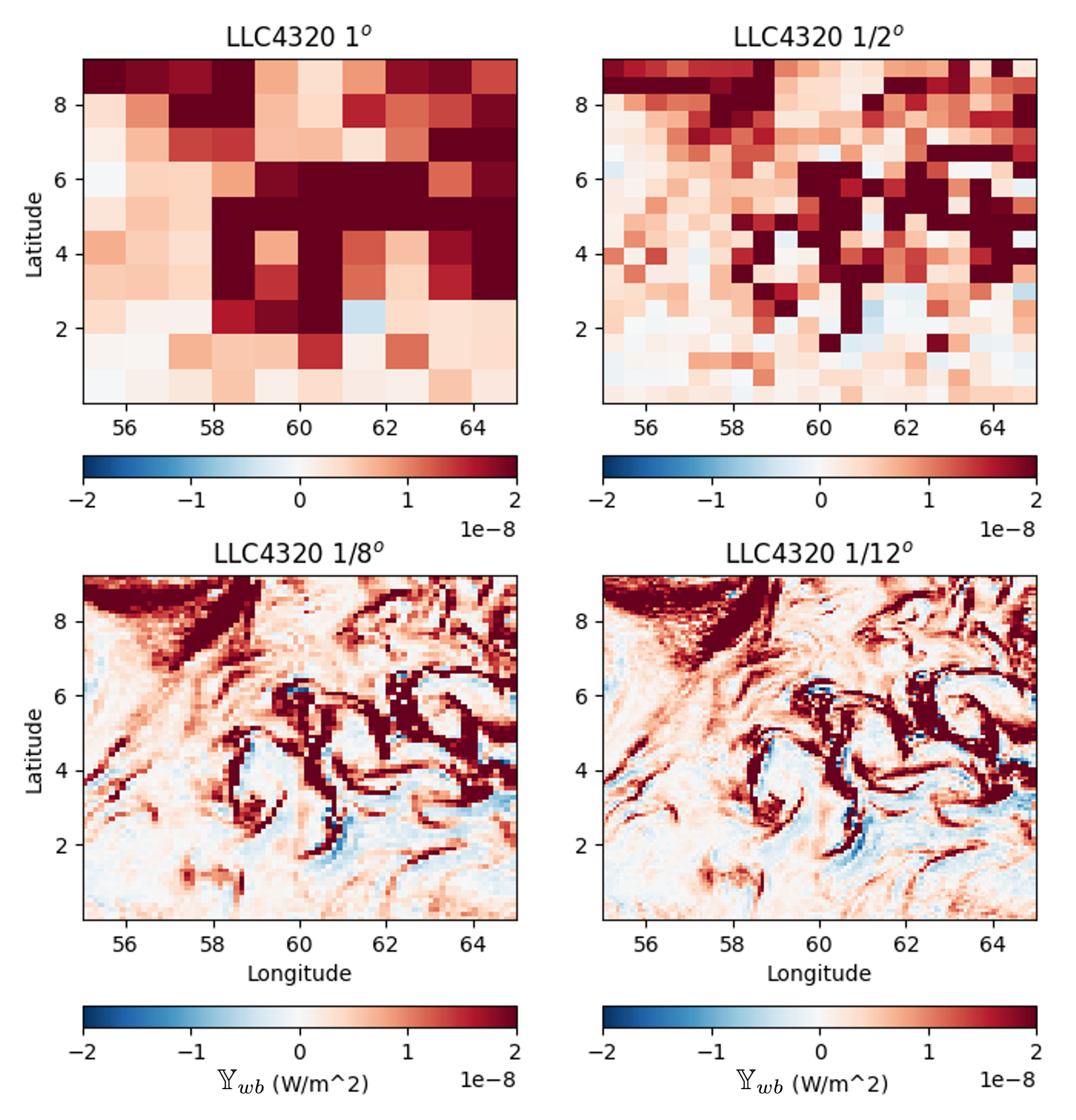}
    \caption{A snapshot as in Figure \ref{fig: predictions snapshot}a to illustrate the CNN output $\mathbb{Y}_{wb}$ (W/m$^2$) with filter scales of $1^o,1/2^o,1/8^o,1/12^o$. } 
  \label{fig: snapshot resolution}
\end{figure}

\begin{figure}[ht!]
 \centering
   \includegraphics[width=1.1\textwidth]{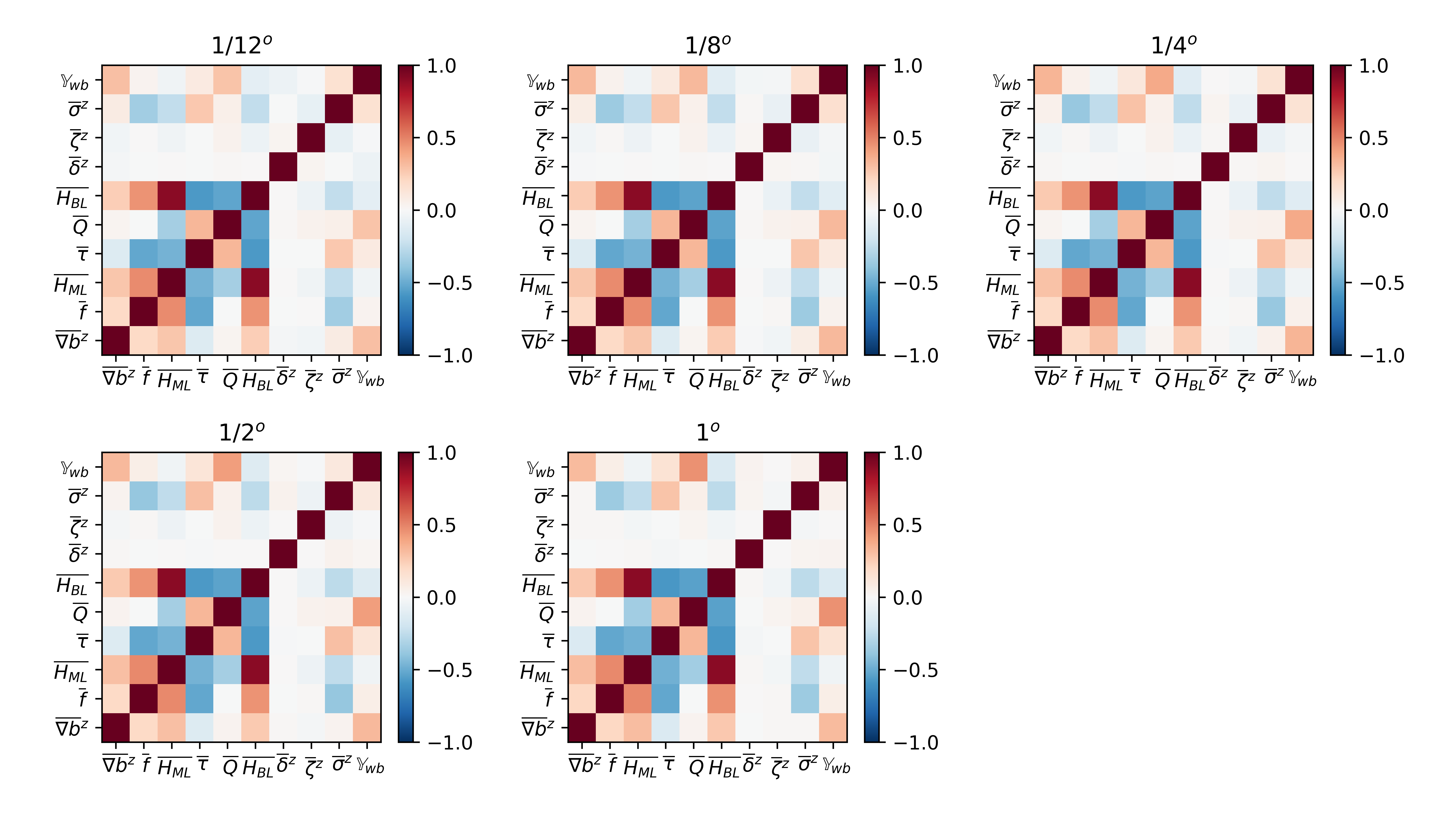}
    \caption{Correlation maps between all input features and output (corresponding to Table \ref{tab:features}) over the entire dataset used for training and testing. Examples of each filter scale applied to data are shown for comparison: $1^o,\ 1/2^o,\ 1/4^o,\ 1/8^o,\ 1/12^o$. Variables associated with surface forcing (e.g., $\overline{H_{BL}}, \overline{Q}, \overline{\tau}$ demonstrate the largest correlation and anti-correlation with each other, dependent on the sign convention. The CNN output, $\mathbb{Y}_{wb}$ shows no correlations above $0.2$ with any of the input features.} 
  \label{fig: corr map}
\end{figure}

\begin{figure}[ht!]
 \centering
   \includegraphics[width=.9\textwidth]{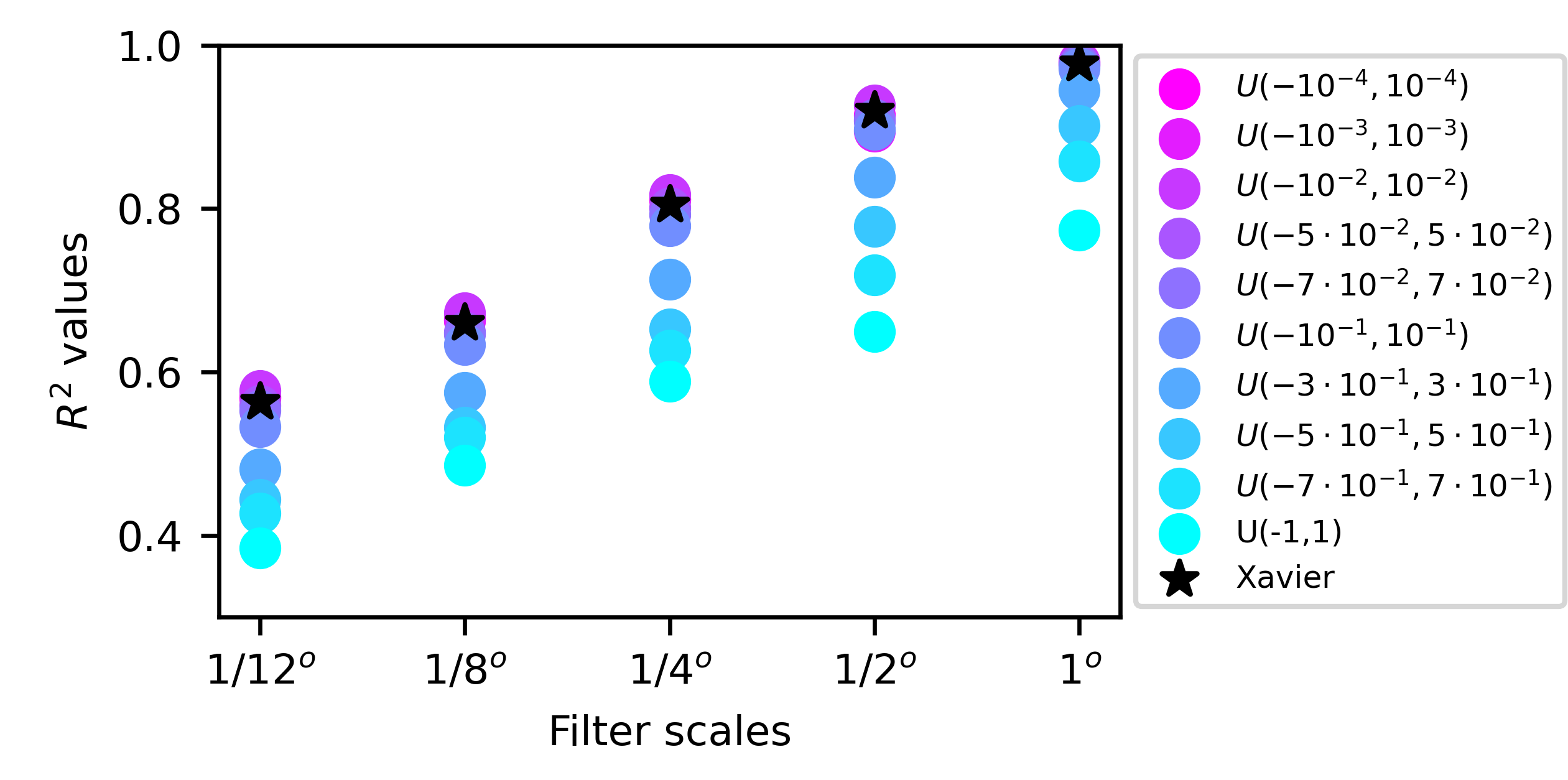}
    \caption{Uniform distributions for weight initializations (noted in the legend by the upper and lower bounds) against the default Xavier initialization in Pytorch, which is a function of the number of input channels. It is shown that the $R^2$ values computed over the entire unseen dataset is  sensitive to the choice of weight initialization. However, the spread in $R^2$ from different initializations is small compared to the different sensitivity tests performed in the main text as long as the distribution bounds stay below $0.1$.} 
  \label{fig: init sensitivity}
\end{figure}

\end{document}